\documentclass[prb,aps,showpacs,twocolumn,preprintnumbers, amsmath,amssymb,superscriptaddress]{revtex4-2}

\usepackage[english]{babel}
\usepackage{amsmath,amssymb,amsfonts}
\usepackage{graphicx}
\usepackage[colorlinks=True,linkcolor=red,citecolor=blue,urlcolor=blue]{hyperref}

\usepackage[dvipsnames]{xcolor}
\usepackage{verbatim}
\usepackage{braket}
\usepackage{bm}
\usepackage{bbm}
\usepackage{braket}
\usepackage{float}
\usepackage{multirow}
\usepackage[normalem]{ulem}
\usepackage{array}
\usepackage{makecell}
\usepackage{subfigure}

\usepackage{xcolor}
\newcolumntype{C}{>{$}c<{$}}

\allowdisplaybreaks
\DeclareMathAlphabet{\zc}{OT1}{pzc}{m}{it}

\def\ket#1{|#1\rangle }

  \DeclareUnicodeCharacter{2212}{-}

\def\changes{\textcolor{black}}

\begin{document}
\title{Percolation-induced PT symmetry breaking}
\author{Mengjie Yang}
\affiliation{Department of Physics, National University of Singapore, Singapore 117551, Singapore}

\author{Ching Hua Lee}
\email{phylch@nus.edu.sg}
\affiliation{Department of Physics, National University of Singapore, Singapore 117551, Singapore}

\date{\today}

\begin{abstract}
We propose a new avenue in which percolation, which has been much associated with critical phase transitions, can also dictate the asymptotic dynamics of non-Hermitian systems by breaking PT symmetry. Central to it is our newly-designed mechanism of topologically guided gain, where chiral edge wavepackets in a topological system experience non-Hermitian gain or loss based on how they are topologically steered. For sufficiently wide topological islands,  this leads to irreversible growth due to positive feedback from interlayer tunneling. As such, a percolation transition that merges small topological islands into larger ones also drives the edge spectrum across a  real to complex transition. Our discovery showcases intriguing dynamical consequences from the triple interplay of chiral topology, directed gain and interlayer tunneling, and suggests new routes for the topology to be harnessed in the control of feedback systems.
\end{abstract}

\maketitle
\newpage


In a non-Hermitian system, the reality of the spectrum crucially controls the long-time dynamics. Even under strong gain or loss, non-Hermitian real spectra and non-divergent time evolution are still possible if PT symmetry remains unbroken~\cite{bender1998real,bender2007making,bender2023pt}. This has been extensively studied and achieved through careful control of gain/loss, coupling strengths and external applied fields~\cite{el2007theory,makris2008beam,longhi2009bloch,bendix2009exponentially,ruter2010observation,lin2011unidirectional,schindler2011experimental,mostafazadeh2013invisibility,zhu2013one,zyablovsky2014pt,chang2014parity,sounas2015unidirectional,konotop2016nonlinear,liu2016metrology,assawaworrarit2017robust,el2018non,chen2018generalized,lee2022exceptional,liu2018observation,ozdemir2019parity,shao2020non}. In optics, for instance, tuning the gain-loss contrast or the refractive index can result in unidirectional invisibility~\cite{lin2011unidirectional,mostafazadeh2013invisibility,zhu2013one,sounas2015unidirectional}, power oscillations~\cite{makris2008beam,longhi2009bloch,bendix2009exponentially,ruter2010observation,schindler2011experimental}, non-reciprocal transmission~\cite{chang2014parity,liu2018observation,shao2020non,ozdemir2019parity} and enhanced sensing~\cite{liu2016metrology,el2018non,chen2018generalized,ozdemir2019parity}.

In this work, we unveil a new route for PT symmetry breaking where the asymptotic dynamics in a disordered non-Hermitian system is controlled by the percolation transition of its real-space texture. While percolation has been extensively linked to second-order phase transitions~\cite{stauffer1979scaling,essam1980percolation,hu1984percolation,parshani2010interdependent,lemoult2016directed,di2024percolation}, conformal field theory~\cite{cardy1992critical,mathieu2007percolation,mertens2012continuum,javerzat2023evidences}, Schramm-Loewner Evolutions~\cite{li2007correlations,mavko1997effect,coniglio1982cluster,balberg1984percolation,javerzat2024schramm} and other critical phenomena~\cite{essam1980percolation,hinrichsen2000non,goltsev2006k,dorogovtsev2008critical,kim2023continuous,muller2023significance,wang2023collective,das2023critical,meng2021concurrence,deger2022constrained,krishnaraj2020coherent,falsi2021direct,wang2020two,lavasani2021topological}, \changes{percolation itself has never been found to control long-time non-Hermitian dynamics~\cite{shante1971introduction,kirkpatrick1973percolation,bollobas2006percolation,saberi2015recent}.} 


Central to our percolation-induced PT symmetry breaking is the newly-designed mechanism of \emph{topologically guided} gain/loss. Unlike in usual non-Hermitian skin effect (NHSE)~\cite{lee2016anomalous,alvarez2018non,yao2018edge,kunst2018biorthogonal,lee2019anatomy,okuma2020topological,lin2023topological,okuma2023non,longhi2019topological,kawabata2019symmetry,lee2019hybrid,song2019non,helbig2020generalized,li2020critical,xue2022non,lee2020unraveling,zou2021observation,zhang2021observation,yang2022concentrated,li2022non,longhi2022self,gu2022transient,shen2022non,jiang2023dimensional,tai2023zoology,longhi2020non,lin2022observation} systems where wavepackets simple spread out and are amplified asymmetrically due to directed amplification, wavepackets in our protocol are \emph{steered} by chiral topological pumping which controls the extent of gain/loss. By weakly coupling two Chern topological layers with oppositely directed gain, this topological guiding mechanism can induce a PT transition when a sufficiently wide distance is traversed -- in the disordered context, this leads to a percolation-induced PT transition as small topological islands coalesce into larger ones. Through this, percolation thus becomes a control knob for the PT transition from oscillatory to divergent dynamics. We note that our mechanism, which relies on chiral topologically guided gain, is completely distinct from the well-known complex-real spectral transition induced by the NHSE~\cite{lee2019anatomy,okuma2020topological,lin2023topological,okuma2023non}, where a boundary blocks directionally pumped state from further gain. \changes{It also contrasts with related works on disordered non-Hermitian systems~\cite{claes2021skin,lin2022observation,zhang2023bulk} which primarily focus on Anderson localization, instead of active directed amplification}.

\begin{figure}
    \centering
    \includegraphics[width=0.5\textwidth]{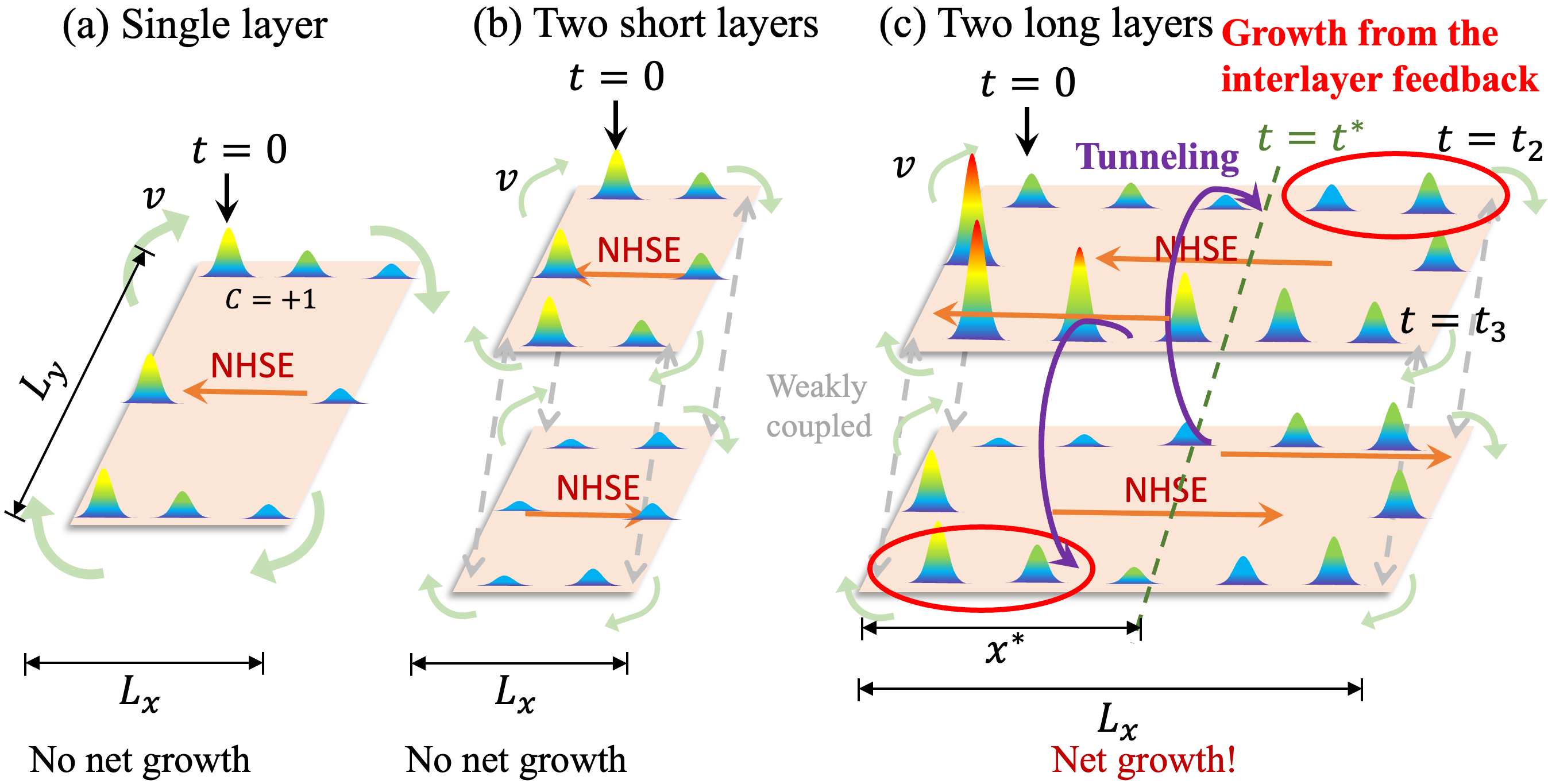}
    \caption{Schematic of the general mechanism of topologically guided gain, which can lead to unlimited growth of edge states when the system is sufficiently large. (a) In a single topological layer, a chiral edge wavepacket encircles a rectangular island boundary with equal attenuation and growth as it propagates against and along the direction of the gain, and thus experiences no net growth. (b) Two sufficiently short weakly coupled layers with oppositely directed gain. After a whole cycle, there is insufficient interlayer feedback~\cite{suppmat}, and thus still no net growth. (c) Two layers with opposite gain directions and sufficiently large width $L_x$. Due to interlayer tunneling feedback, the state starts to grow after propagating longer than a distance $x^*$ along an edge. Hence it experiences net growth approximately given by Eq.~\ref{ImE}.
        }
    \label{fig:fig1}
\end{figure}

\noindent \textit{Size-dependent PT edge transition from topologically guided gain.-- } To understand how a percolation transition can break PT symmetry, we first examine how a single topological island can exhibit PT symmetry breaking when it becomes sufficiently wide. For simplicity, consider an island defined by a rectangular lattice with open boundaries (Fig.~\ref{fig:fig1}). In emphasizing the generic nature of topologically guided gain, we shall not specialize to any particular model yet but only stipulate the following 3 necessary ingredients: (i) chiral topological edge modes of velocity $v$, (ii) directed amplification of strength $\kappa$, and (iii) presence of weak interlayer tunneling rate $\mu$ in a minimal bilayer structure. In Fig.~\ref{fig:fig1}, we depict how the dynamical evolution of edge wavepackets, initialized at time $t=0$ in the upper left corner, is controlled by the triple interplay of these ingredients.

To build intuition, we first consider only a single Chern layer of width $L_x$ and height $L_y$ (Fig.~\ref{fig:fig1}(a)), which harbor the first two ingredients of Chern topology and asymmetric gain/loss. 
An edge wavepacket is chirally pumped clockwise along the rectangular boundaries at a constant speed $v$, remaining confined and preserving its shape even after completing a boundary loop~\cite{bernevig2013topological,qi2011topological,hasan2010colloquium,asboth2016short}. If the lattice hoppings~\cite{suppmat} were made uniformly asymmetric in one direction \changes{i.e. horizontally, without loss of generality}, this edge circulation still remains unitary despite the non-Hermiticity, since the hopping asymmetry can be ``gauged out'' by redefining the real-space basis~\cite{el2018non,yao2018edge,lee2019anatomy,lee2020unraveling}. If the left hoppings $\propto c^\dagger_{x-1}c_x$ are stronger than the right hoppings $\propto c^\dagger_{x+1}c_x$ by a factor of $e^{2\kappa}$, the asymmetry can be removed by rescaling the basis orbital at site $x$ by $c^\dagger_x \rightarrow e^{-\kappa x} c^\dagger$, $c_x\rightarrow e^{\kappa x}c_x$. Hence the wavepacket is simply attenuated by $e^{-\kappa L_x}$ as it is chirally propagated to the right boundary, and grows back to its original magnitude when it returns to the left boundary. Overall, there is thus no net gain after each chiral loop around a single layer. 

We next introduce the third ingredient of weak interlayer tunneling $\mu$ and show how it can lead to topologically guided gain. We weakly couple two clockwise chiral Chern layers with oppositely hopping asymmetries $\pm\kappa$ (Fig.~\ref{fig:fig1}(b,c)). For very weak $\mu$, wavepackets continue to evolve independently in each layer, as in Fig.~\ref{fig:fig1}(b). 

But interestingly, when $\mu$ or the island width $L_x$ is sufficiently large~\footnote{But still with $\mu\ll \mathcal{O}(1)$ much smaller than the topological gap.}, positive feedback growth is possible due to interlayer tunneling. In Fig.~\ref{fig:fig1}(c), the initial wavepacket  $\psi(t=0)=(\psi^U,\psi^L)^T(t=0)=(\psi^U_0,0)^T$ only has nonzero amplitude $\psi^U_0$ in the upper layer. \changes{Due to interlayer tunneling, it quickly induces a small amplitude $\psi^L\sim \mu\psi^U_0$ in the lower layer~\footnote{Here, we assume that the interlayer coupling acts linearly for simplicity. But as shown in~\cite{suppmat}, topologically guided gain remains robust even with significant nonlinearity.}}. Even though $\psi^L$ may initially be very weak, it chirally propagates to the right and is thus amplified exponentially viz. $\psi^L(t)\sim e^{\kappa x}\mu\psi^U_0=e^{\kappa vt}\mu\psi^U_0$. By contrast, $\psi^U(t) \sim e^{-\kappa vt}\psi^U_0$ decays exponentially in the upper layer due to its opposite $\kappa$. For sufficiently large $\mu$ or $L_x$, there would exist a point $x^*=vt^*<L_x$ where $\psi^U$ has decayed so much that it becomes dominated by the tunneling from the growing $\psi^L$: this occurs when $\psi^U(t^*)\sim\mu\psi^L(t^*)$ i.e.
\changes{
\begin{equation}
x^*=vt^*\approx -\kappa^{-1}\log\mu
\label{xcrit}
\end{equation}
for sufficiently weak tunneling $\mu<1$, beyond which gain will trivially occur without system size dependence.
}


After $t=t^*$, wavepackets in \emph{both} layers universally experience exponential gain since the growth is now dominated by the growing lower-layer wavepacket $\psi^L$. The gain stops only at $t_2=L_x/v$ when both wavepackets have arrived at the right boundary, where $\psi^U(t_2)\sim \mu^2 e^{\kappa L_x}\psi^U_0<\psi^L(t_2)\sim \mu e^{\kappa L_x}\psi^U_0$. After chiral propagating to the lower right corner at $t=t_3\approx (L_x+L_y)/v$, the above process repeats with the roles of the upper and lower layer wavepackets reversed, leading to another stationary point in the wavepacket decay (now of $\psi_L$) at $t=t_3+t^*$. In all, after completing one loop around the island boundary, we have $\psi^U\rightarrow \psi^U\mu^2e^{2\kappa L_x}=\psi^U e^{2\kappa (L_x-x^*)}$, which is a net gain. Modeled as non-Hermitian stroboscopic time evolution $\psi(t) =e^{iEt}\psi(0)\sim e^{(\text{Im}E)t}\psi(0)$, where each period lasts for a time interval of $T_\text{period}=2t_3$, this gain corresponds to a complex energy $E$ with 
\begin{equation}
\text{Im}[E]=\frac{2\kappa (L_x-x^*)}{T_\text{period}}=v\frac{\kappa(L_x-x^*)}{L_x+L_y}
\label{ImE}
\end{equation}
when $L_x>x^*$, and zero otherwise. \changes{Note that this size dependence only depends on the number of unit cells and the hopping asymmetry between them, not the physical length of the system.}

Just from purely heuristic arguments, independent of model details, we have managed to show how the triple interplay of chiral topology, directed gain and interlayer tunneling can break edge PT symmetry proportionally to the velocity $v$ of the chiral edge modes, and the gain asymmetry $\kappa$. Remarkably, the dynamical gain rate $\text{Im}[E]$ is also strongly dependent on the island width and aspect ratio, although in a manner distinct from the critical skin effect~\cite{li2020critical,liu2020helical,yokomizo2021scaling,qin2023universal}. \textcolor{black}{Indeed, while the physics of topological guided gain may draw parallels with a quasi-1D system experiencing critical NHSE, the latter crucially does not involve edge-localized chiral topological pumping in controlling wavepackets dynamics~\footnote{In the the critical skin effect, the exponential growth across directed hoppings occurs with unguided spreading of the state, rather than along a well-defined chiral edge trajectory.}}. 


\begin{figure*}[!htp]
    \centering
   \includegraphics[width=0.85\textwidth]{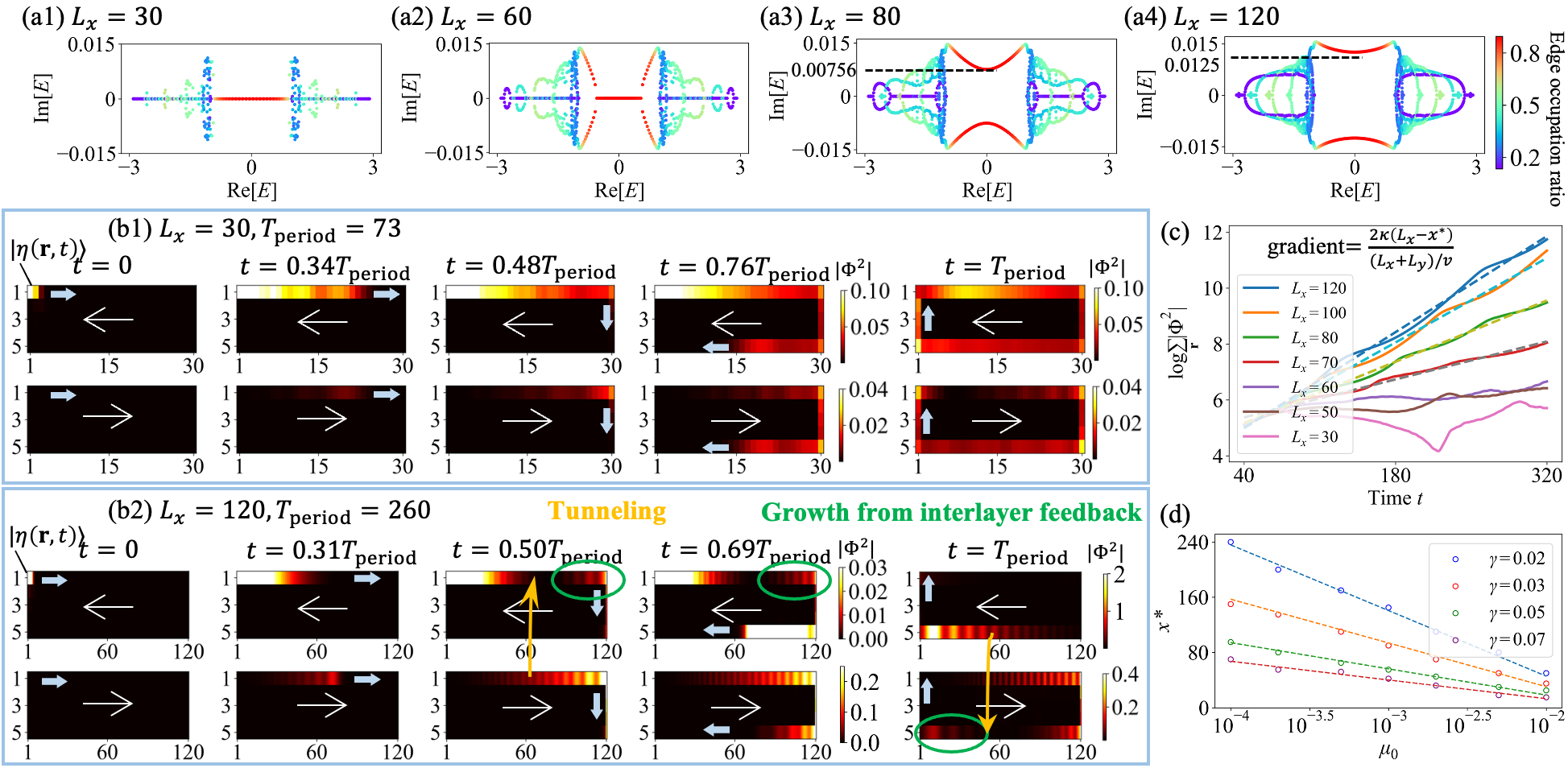}
    \caption{Edge PT breaking from topologically guided gain in rectangular $L_x\times L_y$ $\hat H_\text{two-layer}$ islands. (a) Spectral plots reveal that edge eigenenergies (colored red via their edge occupation ratio Eq.~\ref{eq:ratio}) transition from real to complex as $L_x$ exceeds the critical width of $x^*\approx 60$, for $\gamma=0.02, \mu_0=0.01$. 
(b) Snapshots of the dynamical state evolution $\Phi(\bold r, t)$ from an excitation $\eta(\bold r,t)$ at the upper left corner of the upper layer, for 
(b1) short and (b2) long $L_x$, with fixed $L_y=5$. White arrows indicate the oppositely directed gains in the upper and lower layers, while pale blue arrows indicate the universal clockwise topological pumping. Consistent with the picture from Figs.~\ref{fig:fig1}(b) and (c) respectively, for $L_x<x^*$ (b1), topological edge modes circulate the upper layer's perimeter unaided by the relatively weak $\Phi(\bold r,t)$ in the lower layer; but for $L_x>x^*$ (b2), states that tunneled to the lower layer grows via $\sim e^{\kappa x}$ for a longer distance $x$, eventually reaching the $x>x^*$ region where it induces growth in the upper layer edge, leading to topological guided gain.
(c) The evolution of the integrated state density $\sum_\bold r|\Phi(\bold r,t)|^2$ with time, showing exponential growth when $L_x>x^*$. 
The growth rate from the gradient (Eq.~\ref{MaxImE}) agrees with the topological guided gain mechanism (Eq.~\ref{ImE}), with dynamically determined chiral velocity $v=0.96$ and $\kappa=0.024$.
(d) Numerically measured critical transition widths $x^*$ at different hopping asymmetries $\gamma$, which is logarithmically decaying with interlayer coupling $\mu_0$, consistent with Eq.~\ref{xcrit} with $\mu\approx 33\mu_0$ and hopping asymmetry $\kappa=\ln\left|\sqrt{\frac{m+\cos k_y+\gamma}{m+\cos k_y-\gamma}}\right|$ with $k_y\approx 0.56\pi$ and $m=1$.
}
\label{fig:fig2}
\end{figure*}

For a concrete demonstration of topologically guided gain, we next specialize in a coupled bilayer system where each layer is given by the well-known QWZ Chern model~\cite{qi2006topological} with non-Hermitian asymmetry $\gamma$ introduced into every other hopping in the $x$-direction~\cite{kawabata2018anomalous}:
\begin{equation}
    \begin{aligned}
        \hat{H}_{\text{two-layer}}(\boldsymbol{k},\gamma,\mu) &= \begin{pmatrix}
            \hat{H}_\text{Ch}(\boldsymbol{k},\gamma) & \mu_0\,\mathbb{I} \\
            \mu_0\,\mathbb{I} & \hat{H}_\text{Ch}(\boldsymbol{k},-\gamma)
        \end{pmatrix},\\
        H_\text{Ch}(\boldsymbol{k},\gamma) &= (m +  \cos k_x +  \cos k_y) \sigma_x \\
        &\quad + (\mathrm{i} \gamma + \sin k_y) \sigma_y + ( \sin k_x) \sigma_z
    \end{aligned}
    \label{eq:twolayer}
\end{equation}
where $m=1$ such that the Chern number is $+1$, and $\sigma_x, \sigma_y$ and $\sigma_z$ the Pauli matrices in the sublattice basis.
The interlayer coupling $\mu_0$ is expected to be proportional to the interlayer tunneling rate $\mu$, which we shall later determine empirically. 
With $\gamma>0$, the directed gain is towards the left(right) in the upper(lower) layer. Confined by boundaries in both directions under a $L_x\times L_y$ rectangular geometry, this directed gain modifies the energy dispersion to $E^2=(m+\cos k_y +\gamma+z^{-1})(m+\cos k_y-\gamma+z)+\sin^2 k_y $, with $z=e^{-\kappa}e^{ik_y}$, $\kappa=-\log\sqrt{|(m+\cos k_y-\gamma)/(m+\cos k_y+\gamma)|}$ as determined by the generalized Brillouin zone~\cite{suppmat,kawabata2018anomalous,lee2019anatomy}.

Most saliently, $\hat{H}_{\text{two-layer}}$ exhibits real-to-complex edge PT breaking as the system width $L_x$ is increased beyond a certain critical $x^*$ (with fixed $L_y=5$ unit cells), as predicted by Eq.~\ref{ImE} and confirmed in the spectral plots of Fig.~\ref{fig:fig2}(a1)-(a4). Here $x^*\approx 60$ unit cells, as confirmed by detailed dynamical simulations below. Each eigenenergy $E$ in Fig.~\ref{fig:fig2}(a) is colored according to the edge occupation ratio of its eigenstate $\psi(\boldsymbol{r})$, defined as 
\begin{equation}
    \text{edge occupation ratio}=\frac{\sum_{\boldsymbol{r}\in \text{edges}}|\psi(\boldsymbol{r})|^2}{\sum_{\boldsymbol{r}}|\psi(\boldsymbol{r})|^2},\label{eq:ratio}
\end{equation}
such that the bulk spectra appear cyan or blue while the edge modes are red. Evidently, edge eigenenergies (red) transition from real to complex as $L_x$ increases, becoming entirely complex by $L_x=80$.

\begin{figure*}[!htp]
    \centering
    \includegraphics[width=0.85\textwidth]{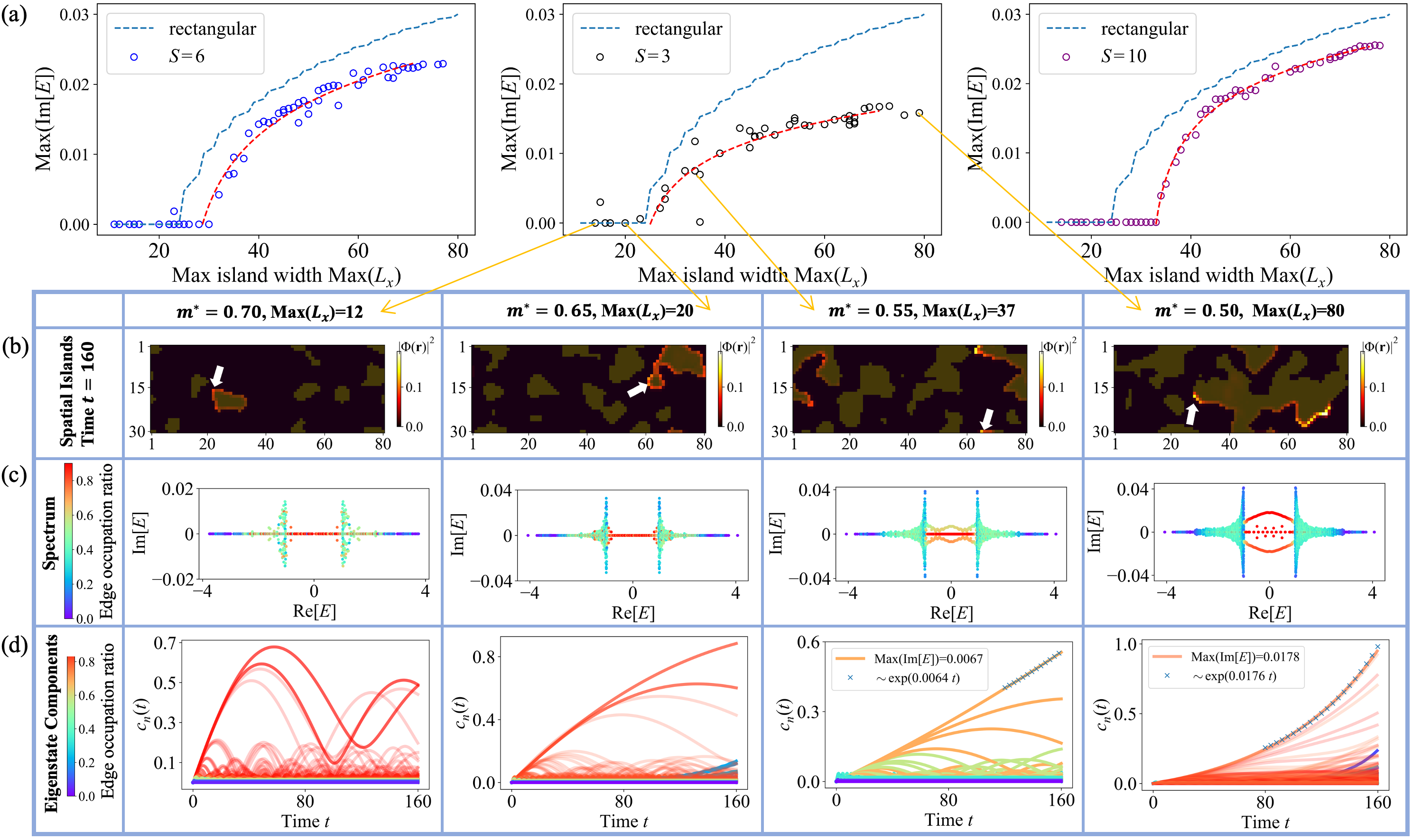}
    \caption{Percolation-induced edge PT transition from topologically guided gain in $\hat{H}_{\text{two-layer}}$ with $\mu_0=0.005,\gamma=0.05$. (a) As topological islands coalesce, the largest width $\text{Max}(L_x)$ among them increases, eventually resulting in a real-to-complex edge spectral transition, as captured by the abrupt jump in $\text{Max}(\text{Im}[E])$ for islands of various boundary smoothness $S$. 
(b) Illustrative island configurations and dynamics for $S=3$,	with $\text{Max}(L_x)$ increasing as islands coalesce due to the dropping ``sea level'' $m^*$. Shown are the $|\Phi(\bold r,t=160)|^2$ snapshots from edge drivings $\eta(\bold r,t)$ (white arrows), where greater gain (yellow) is manifestly observed around wider islands. 
(c) Their corresponding spectra, with edge states (colored orange or red from Eq.~\ref{eq:ratio}) become complex as $m^*$ decreases. 
(d) Evolution of eigenbasis expansion coefficients $c_n(t)$ of the evolved state $\Phi(\bold r,t)$, with growth rate accurately agreeing with $\text{Max}(\text{Im}[E])$ for clean edge propagation ($m^*=0.50$), and qualitative agreeing when some bulk leakage is present ($m^*=0.55$). For larger $m^*$ with $\text{Max}(\text{Im}[E])=0$, $c_n(t)$ oscillates until significant bulk leakage occurs.
		}
    \label{fig:fig3}
\end{figure*}

To definitively show that this edge PT transition indeed occurs due to topological guided gain mechanism, 
we simulate the state dynamics in $\hat{H}_{\text{two-layer}}$ due to a source excitation $\ket{\eta(\bold r,t)}=\eta_0e^{-i\omega_s t}|\bold r_s\rangle$ at $\bold r= \bold r_s$, the top left corner of the upper (but not lower) layer. To resonantly excite the topological edge modes~\cite{suppmat} and minimize bulk leakage,
we simulated the dynamics using the frequency-shifted Hamiltonian $\hat{H}^\prime_{\text{two-layer}}=\hat{H}_{\text{two-layer}}+\omega_s\mathbb{I}$, such that $E'=\omega_s$ lies exactly in the middle of the topological gap. We set $\omega_s=2$ and $\eta_0=0.1$. 

Fig.~\ref{fig:fig2}(b) showcases the dynamical state evolution $\ket{\Phi(\bold r,t)}$ to the time-dependent differential equation $\ket{\dot{\Phi}(\bold r,t)}=-i\hat{H}^\prime_{\text{two-layer}(\bold r)}\ket{\Phi(\bold r,t)}+\ket{\eta(\bold r,t)}$ over one loop $T_\text{period}$, for short and long $\hat{H}_{\text{two-layer}}$ lattices ($L_x=30$ and $120$), corresponding to the real and complex edge spectra in Figs.~\ref{fig:fig2}(a1) and (a4) respectively. For short $L_x=30<x^*$, the state shrinks as it is chirally transported towards the right in the upper layer and grows back as it loops back, hardly affected by the lower layer, reminiscent of Fig.~\ref{fig:fig1}(b). But for long $L_x=120>x^*$, the state grows simultaneously in both layers in the region $x>x^*=60$, thereby giving rise to net gain, as also foreshadowed in Fig.~\ref{fig:fig1}(c). \changes{These heuristics, for sufficiently small $\mu$, agree with quantitative predictions by an effective model in the subspace of topological edge states~\cite{suppmat}.}

To more clearly observe that $L_x>x^*$ leads to net gain, we plot the numerically obtained total state amplitude $\sum_{\bold r}\langle\Phi(\bold r,t)|\Phi(\bold r,t)\rangle=\sum_{\boldsymbol{r}}|\Phi(\bold r,t)|^2$ over the whole system in Fig.~\ref{fig:fig2}(c). For $L_x< x^*$, there is no sustained gain at long times, as expected from the real edge spectra. But for $L_x>x^*$, the empirical topologically guided gain~\cite{suppmat}
\begin{equation}
\text{Max}\left(\text{Im}[E]\right)\approx \frac1{2}\frac{d}{dt}\log\sum_r|\Phi(\bold r,t)|^2
\label{MaxImE}
\end{equation}
can be read from the nonzero asymptotic gradient of the log plot. It exhibits close agreement with the growth rate predicted by Eq.~\ref{ImE} (dashed) using the dynamically determined values of the chiral velocity $v=0.96$ and inverse spatial decay length $\kappa=0.024$ (with $k_y \approx 0.56\pi$)~\cite{suppmat}.

The soundness of our theoretical framework is further verified in Fig.~\ref{fig:fig2}(d), where $x^*$ is shown to decrease logarithmically with the interlayer coupling $\mu_0$ across various non-Hermiticity $\gamma$, exactly consistent with Eq.~\ref{xcrit} $x^* =-\kappa^{-1}\log\mu$ (dashed) \changes{if we set $\mu= 33\mu_0$, the $\kappa$-dependent constant of proportionality $33$ obtained via fitting.} Here, the $x^*$ data points are empirically obtained as the smallest $L_x$ whereby $\sum_r|\Phi(\bold r,t)|^2$ starts to increase exponentially with time.

\noindent
\textit{Percolation-induced PT symmetry breaking. --}
Having established that rectangular islands exhibit edge PT breaking due to topologically guided gain when their width $L_x>x^*$, we next show that this applies also to disordered islands that can dramatically grow via percolation. In this work, percolation is effected by adjusting the ``sea level'' $m^*$ upon a scalar landscape $m(\bold r)$ with random Gaussian correlation $\langle m(\bold r_1)m(\bold r_2) \rangle=\langle m^2\rangle e^{-|\boldsymbol{r_1}-\boldsymbol{r_2}|^2/(2S^2)}$, whose generation is detailed in~\cite{suppmat}. Regions $\bold r$ where $m(\bold r)\geq m^*$ are above the ``sea level'' are designated to be nontrivial topological Chern islands ($m=1$ in Eq.~\ref{eq:twolayer}), with boundary smoothness controlled by the correlation length $S$. From previous arguments, the widest island (with width denoted as $\text{Max}(L_x)$) is expected to control the asymptotic state growth.

As illustrated in Fig.~\ref{fig:fig3}(b) in a periodic $80\times 30$-unit cell system with $S=3$, decreasing the sea level $m^*$ allows the topological islands (white) to grow and coalesce, eventually spanning the entire system as $\text{Max}(L_x)$ saturates at $80$. How exactly the distribution of the island sizes depends on $m^*$ is described in~\cite{suppmat}.

Fig.~\ref{fig:fig3}(a) shows the distinctive real-to-complex spectral transition of $\text{Max}(\text{Im}[E])$ of the \emph{edge} modes as the largest island width $\text{Max}(L_x)$ increases. Whether the island boundaries are rough ($S=3$) or smooth ($S=10$), we always have $\text{Max}(\text{Im}[E])>0$ when $\text{Max}(L_x)$ is sufficiently large i.e. where at least one island is wide enough to support topologically guided gain. 
Crucially in this disordered scenario, the edge dynamics involve a competition between topologically guided propagation/gain and bulk leakage, \changes{which can also lead to unlimited growth up to physically-imposed constraints i.e. breakdown of electrical circuit components with excessively strong electric fields or Joule heating~\cite{suppmat}.
}
While small island features can hinder the gain and lead to lower growth ($\text{Im}[E]$) compared to that in perfectly rectangular islands (blue dashed), they also introduce fluctuations in $\text{Max}(\text{Im}[E])$ due to bulk leakage.

Figs.~\ref{fig:fig3}(b)-(d) showcases the spectral and corresponding dynamical behavior in four specific disordered instances with $S=3$, which allows for small island features without excessively sharp boundaries that overshadow the edge PT transition with bulk leakage. For sufficiently large $\text{Max}(\text{Im}[E])=37$ and $80$, the complex edge spectra (red in Figs.~\ref{fig:fig3}(c)) can indeed be attributed to the topologically guided gainy propagation around a sufficiently large island, as displayed in Figs.~\ref{fig:fig3}(b). 

To unambiguously confirm that the numerical state dynamics are indeed governed by $\text{Max}(\text{Im}[E])$ obtained from diagonalization, we further plot in Figs.~\ref{fig:fig3}(d) the eigenbasis expansion coefficients $c_n(t)=\langle\psi_n(\bold r)\ket{\Phi(\bold r,t)}$ of the time-evolved state $\Phi(\bold r,t)$, where $n$ ranges over all $2L_xL_y$ eigenstates. When $\text{Max}(\text{Im}[E])=0$ as for $m^*=0.70$ and $0.65$, the $c_n(t)$ coefficients oscillate without clear exponential growth; while for the $m^*=0.50$ case dominated by a $\text{Max}(L_x)=80$ island stretching across the system edges, the leading $c_n(t)\sim e^{0.0176t}$ from the growth of $\Phi(\bold r,t)$ closely matches the growth rate $\text{Max}(\text{Im}[E])=0.0178$ independently obtained from diagonalization. Topologically guided gain can still remain as the primary evolution mechanism in the presence of some bulk leakage (orange or green), such as in the $\text{Max}(L_x)=37$ case where the leading $c_n(t)$ growth is still qualitatively predicted by $\text{Max}(\text{Im}[E])$.

\noindent
\textit{Conclusion.--} From the triple interplay of chiral topology, directed gain and interlayer tunneling emerges the mechanism of topologically guided gain, where edge wavepackets on sufficiently wide islands experience irreversible growth due to interlayer feedback~\cite{suppmat}, \changes{as quantitatively verified through an effective model~\cite{suppmat}.} We demonstrated this first schematically, then on a single rectangular island, and finally in a disordered landscape of topological islands. For sufficiently smooth islands, a sharp PT symmetry transition occurs as smaller islands percolate and combine to form larger islands beyond a threshold width. 
Moving forward, it would be interesting to explore how topological guiding can interplay with non-linear feedback~\cite{zhou2017optical,smirnova2019topological,tuloup2020nonlinearity,smirnova2020nonlinear,xia2021nonlinear,sone2022topological,hohmann2023observation} and interlayer tunneling subject to Moire interference~\cite{ohta2012evidence,dai2016twisted,kim2017tunable,naik2018ultraflatbands,carr2020electronic,de2022imaging},
as well as experimental demonstrations built upon mature mechanical~\cite{brandenbourger2019non,ghatak2020observation,wen2022unidirectional,xiu2023synthetically}, photonic~\cite{pan2018photonic,xiao2020non,zhu2020photonic,song2020two,ao2020topological}, electrical circuit~\cite{hofmann2019chiral,ezawa2019electric,helbig2020generalized,hofmann2020reciprocal,liu2020gain,liu2021non,stegmaier2021topological,zhang2020non,zhang2022observation,shang2022experimental,yuan2023non,zou2023experimental,zhu2023higher,zhang2023electrical,suppmat} and quantum circuit platforms~\cite{smith2019simulating,gou2020tunable,koh2022simulation,kirmani2022probing,frey2022realization,chertkov2023characterizing,chen2023robust,yang2023simulating,iqbal2023creation,shen2023observation,koh2023observation}. \changes{A detailed experimental proposal based on a topolectrical circuit array is provided in the Supplemental Material~\cite{suppmat}.}

\bibliography{ref_final.bib}


\onecolumngrid
\flushbottom
\newpage

\setcounter{equation}{0}
\setcounter{figure}{0}
\setcounter{table}{0}
\setcounter{section}{0}
\setcounter{page}{1}
\renewcommand{\theequation}{S\arabic{equation}}
\renewcommand{\thefigure}{S\arabic{figure}}
\renewcommand{\thesection}{S\arabic{section}}
\renewcommand{\thepage}{S\arabic{page}}

\large
\begin{center}
    \textbf{Supplemental Material for ``Percolation-induced PT symmetry breaking''}
\end{center}

\normalsize

\section{Detailed analysis of the dynamical driving term}

In the main text, we have used the fact that the driving term does not affect the state growth rate much when the system is in the exponential growth regime, despite still functioning as an energy source. Here we provide a detailed analysis.


\noindent The Schr\"{o}dinger equation for the state evolution subject to a driving term $|\eta(\bold r,t)\rangle=\eta_0e^{-i\omega_s t}|\bold r_s\rangle$ is 
\begin{equation}
\ket{\dot{\Phi}(t)}=-i\hat{H}^\prime_{\text{two-layer}}\ket{\Phi(t)}+\ket{\eta(\mathbf r,t)},
\end{equation}
where $\hat{H}^\prime_{\text{two-layer}}=\hat{H}_{\text{two-layer}}+\omega_s \hat{I}$ is the Hamiltonian that is frequency-shifted such that the midgap edge mode correspond to nonzero frequency $\omega_s\in \mathbb{R}$. Its general solution is given by 
\begin{equation}
    |\Phi(t)\rangle=\hat{U}\left(t, t_0\right)\left|\Phi\left(t_0\right)\right\rangle+\int_{t_0}^t dt^{\prime}\hat{U}\left(t, t^{\prime}\right)\ket{\eta(\mathbf r,t')}. \label{smeq:Phisol}
\end{equation}
$\hat{U}\left(t, t_0\right)=\exp \left(-i \int_{t_0}^t \hat{H}^\prime_{\text{two-layer}}\left(t^{\prime}\right) d t^{\prime}\right)$ is the time-evolution operator that propagates the \emph{undriven} state from time $t_0$ to time $t$. But with external driving $|\eta(\bold r, t')\rangle$, the second term (driving term) in Eq.~\ref{smeq:Phisol} represents the contribution of the external driving throughout the time interval, whose impact is what we are studying. Indeed, the driving term can be thought of as ``continually'' feeding in a new initial state from $t_0$ to $t$. To proceed, we perform the spectral decomposition of the time-evolution operator $\hat{U}\left(t, t^{\prime}\right)$, which is given by
\begin{equation}
  \hat{U}\left(t, t^{\prime}\right)=\sum_n e^{-i E_n\left(t-t^{\prime}\right)}|\psi_n\rangle\langle\psi_n|,\label{smeq:U}
\end{equation}
where $E_n$ and $|\psi_n\rangle$ are the eigenvalues and eigenvectors of the Hamiltonian $\hat{H}'_{\text{two-layer}}$. Since $E_n$ may be complex for non-Hermitian Hamiltonians, we rewrite the eigenvalues as
\begin{equation}
  E_n = \epsilon_n + i\beta_n,
\end{equation}
where $\epsilon_n$ is the oscillation frequency and $\beta_n$ represents growth/decay (positive/negative). The time evolution operator in Eq.~\ref{smeq:U} then can be rewritten as
\begin{equation}
\hat{U}(t,t')\quad =\quad \sum_n e^{-i\epsilon_n(t-t')}e^{\beta_n(t-t')} |\psi_n\rangle\langle\psi_n|. \label{smeq:U2}
\end{equation}
Plugging Eq.~\ref{smeq:U2} into the second term in the integral Eq.~\ref{smeq:Phisol}, we obtain 
\begin{equation}
  \begin{aligned}
  \int_{t_0}^t dt' \hat{U}(t,t') |\eta(\mathbf{r},t')\rangle  &= \int_{t_0}^t dt' \sum_n e^{-i\epsilon_n(t-t')}e^{\beta_n(t-t')} |\psi_n\rangle\langle\psi_n|\eta(\mathbf{r},t')\rangle
  \\&= \sum_n |\psi_n\rangle \bigg(\eta_0 \int_{t_0}^t dt' e^{-i(\epsilon_n - \omega_s)(t - t')} e^{\beta_n(t-t')} \langle \psi_n|\mathbf{r}_s\rangle \bigg).
  \end{aligned}
\end{equation}
The time integral can be evaluated as
\begin{equation}
  \int_{t_0}^t dt' e^{-i(\epsilon_n - \omega_s)(t - t')} e^{\beta_n(t-t')} =\frac{e^{\beta_n(t - t_0)} - e^{-i(\epsilon_n - \omega_s)(t - t_0)}}{-\beta_n - i(\epsilon_n - \omega_s)}.
\end{equation}
From this, we obtain the contribution of the external driving over the time interval as
\begin{equation}
  \int_{t_0}^t dt' \hat{U}(t,t') |\eta(\mathbf{r},t')\rangle = \eta_0\sum_n \langle \psi_n | \mathbf{r}_s\rangle  \frac{e^{\beta_n(t - t_0)} - e^{-i(\epsilon_n - \omega_s)(t - t_0)}}{-\beta_n - i(\epsilon_n - \omega_s)}|\psi_n\rangle \label{smeq:Phisol2}
\end{equation}
where $\beta_n=\text{Im}[E_n]$ and $\epsilon_n=\text{Re}[E_n]$ as previously defined. This result in Eq.~\ref{smeq:Phisol2} can be explained as follows. $\eta_0\langle \psi_n | \mathbf{r}_s\rangle$ represents the coupling strength between the driving field and the $n$-th eigenstate of the system. The driving term excites the eigenstates of the time-evolution operator that have a nonzero overlap with the spatial profile of the driving term, i.e., $\langle \psi_n | \mathbf{r}_s\rangle \neq 0$. This means that the driving term can selectively excite certain eigenstates, depending on the position $\mathbf{r}_s$ of the driving source. In the main text, we choose $\mathbf{r}_s$ to be at the left-top corner of the rectangular lattice. 
This position is chosen because the topological edge state is extensively localized there. Therefore, the driving term can effectively and selectively excite the topological edge state as intended. 

\begin{figure}[!htp]
    \centering
    \subfigure[]{\includegraphics[width=0.18\textwidth]{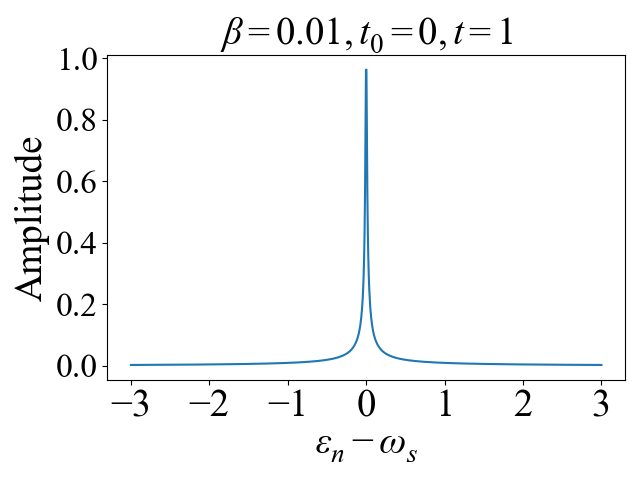}}
    \subfigure[]{\includegraphics[width=0.18\textwidth]{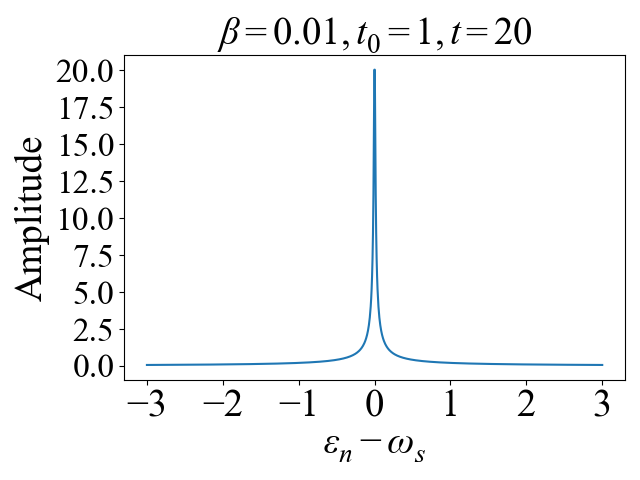}}
    \subfigure[]{\includegraphics[width=0.18\textwidth]{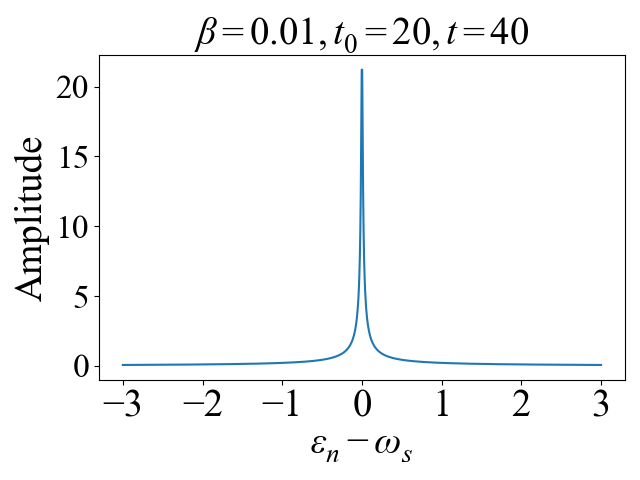}}
    \subfigure[]{\includegraphics[width=0.18\textwidth]{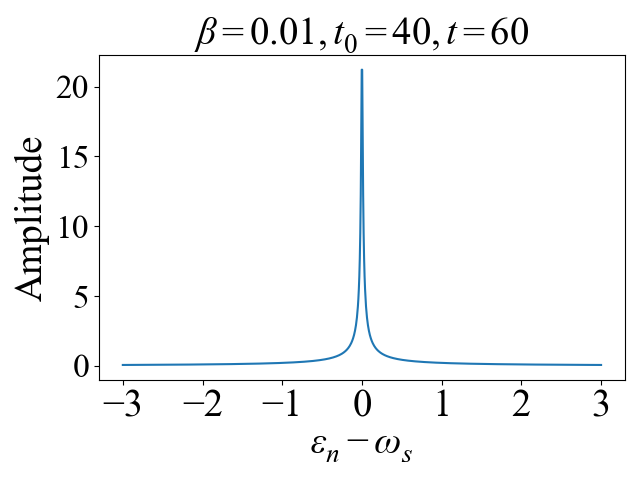}}
    \subfigure[]{\includegraphics[width=0.18\textwidth]{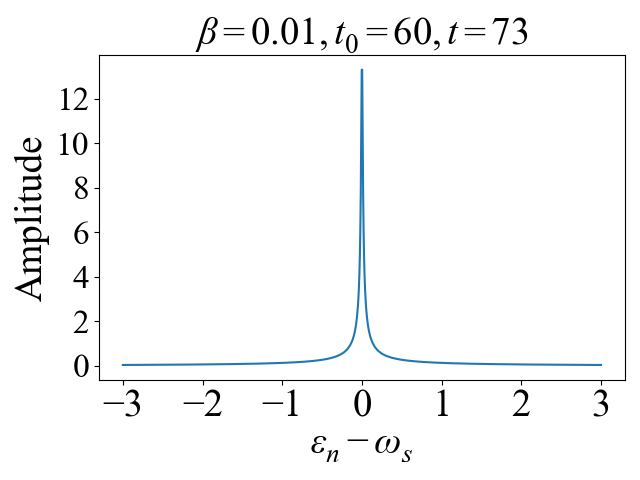}}
    \caption{The absolute value of the complex factor $\left(e^{\beta_n(t - t_0)} - e^{-i(\epsilon_n - \omega_s)(t - t_0)}\right)/\left(-\beta_n - i(\epsilon_n - \omega_s)\right)$ as a function of $\epsilon_n - \omega_s$ for relevant parameters in the main text: $\beta=0.01$, $T_\text{period}=73$ in Fig.~2(b1).
    }
    \label{smfig:why_resonate_73}
\end{figure}

\begin{figure}[!htp]
  \centering
  \subfigure[]{\includegraphics[width=0.18\textwidth]{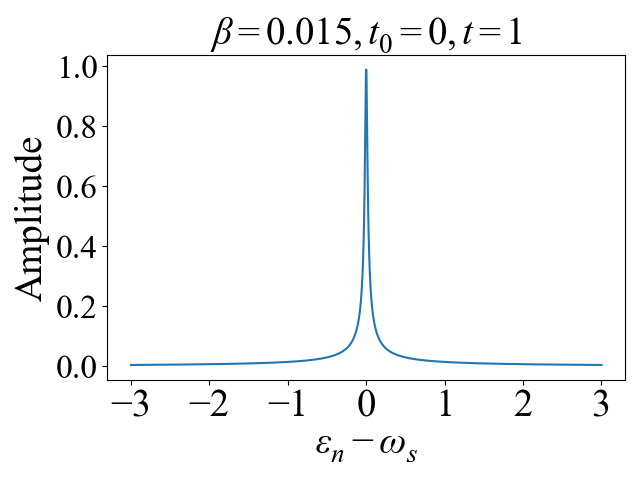}}
  \subfigure[]{\includegraphics[width=0.18\textwidth]{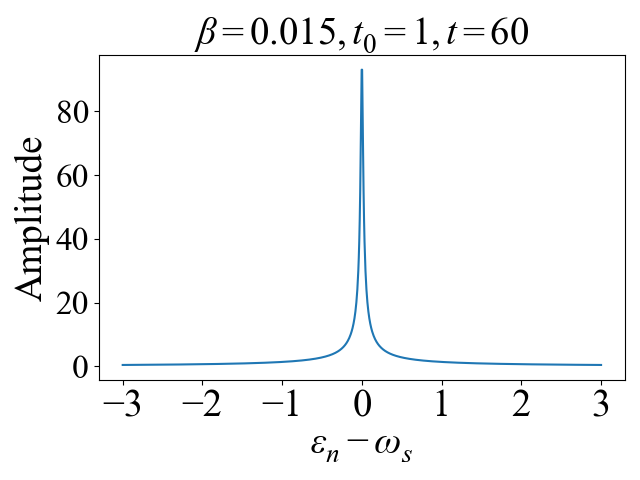}}
  \subfigure[]{\includegraphics[width=0.18\textwidth]{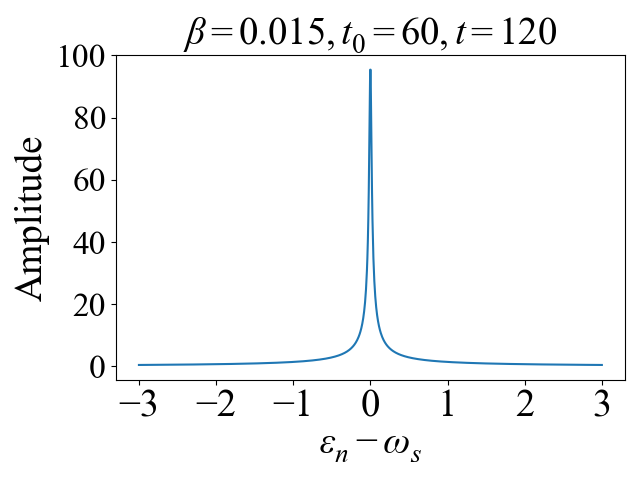}}
  \subfigure[]{\includegraphics[width=0.18\textwidth]{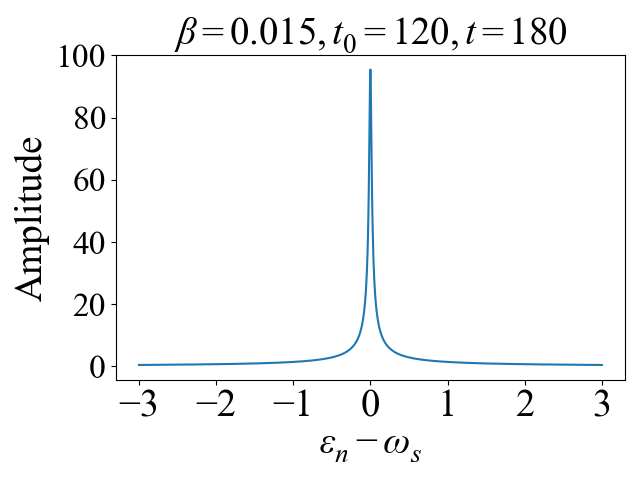}}
  \subfigure[]{\includegraphics[width=0.18\textwidth]{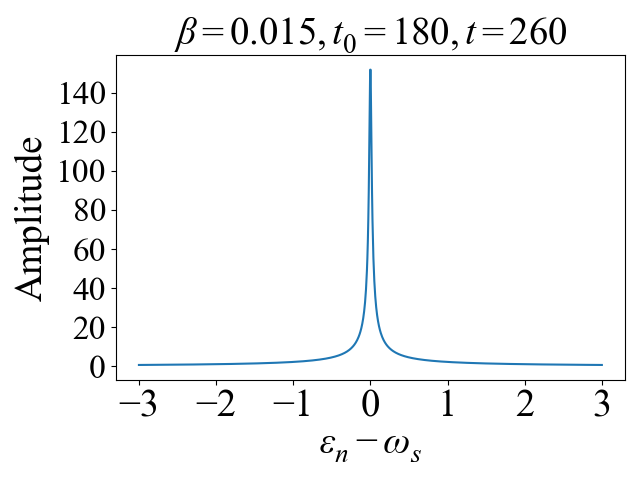}}
  \caption{The absolute value of the complex factor $\left(e^{\beta_n(t - t_0)} - e^{-i(\epsilon_n - \omega_s)(t - t_0)}\right)/\left(-\beta_n - i(\epsilon_n - \omega_s)\right)$ as a function of $\epsilon_n - \omega_s$ for relevant parameters in the main text: $\beta=0.015$, $T_\text{period}=260$ in Fig.~2(b2).
  }
  \label{smfig:why_resonate_260}
\end{figure}

The contribution from the overlap $\langle \psi_n | \mathbf{r}_s\rangle$ is given by the complex factor $\frac{\left(e^{\beta_n(t - t_0)} - e^{-i(\epsilon_n - \omega_s)(t - t_0)}\right)}{\left(-\beta_n - i(\epsilon_n - \omega_s)\right)}$ that depends on the growth rate $\beta_n$, the energy difference $\epsilon_n - \omega_s$, for a time interval $t - t_0$. This factor determines the amplitude and phase of the contribution of the $n$-th eigenstate to the final state of the system. The absolute value of the complex factor is given by
\begin{equation}
  \left|\frac{e^{\beta_n(t - t_0)} - e^{-i(\epsilon_n - \omega_s)(t - t_0)}}{-\beta_n - i(\epsilon_n - \omega_s)}\right| = \frac{\sqrt{e^{2\beta_n(t - t_0)} + 1 - 2e^{\beta_n(t - t_0)}\cos((\epsilon_n - \omega_s)(t - t_0))}}{\sqrt{\beta_n^2 + (\epsilon_n - \omega_s)^2}}.
\end{equation}
Evidently, for a real eigenenergy (such as a PT-unbroken topological edge mode), $\text{Im}[E_n]=\beta_n=0$ and the overwhelming large contribution occurs at resonance $\epsilon_n=\omega_s$. For nonzero $\beta_n$, this contribution grows exponentially, as would an overlap contribution with any initial state. However, at small linewidth $\beta_n$, a broadened resonance peak around $\epsilon_n-\omega_s$ still exists, as illustrated in Figs.~\ref{smfig:why_resonate_73} and \ref{smfig:why_resonate_260}. In the text, we used an $\omega_s$ that resonates with midgap topological edge states $\psi_n$ with zero or very small $\text{Im}[E_n]=\beta_n$ before PT breaking.

\color{black}

\section{The numerical diagonalization of the Hamiltonian and the calculation of Max(Im[$E$])}

In the main text, we relate the growth of $\Phi$ with a value of Max(Im[$E$]), which is the maximum value of the eigenvalues, independently obtained from diagonalization. Below, we elaborate on how these data points are obtained. 

To numerically obtain the value of Max(Im[$E$]), we diagonalize the Hamiltonian and calculate the eigenvalues. The maximum imaginary part of the eigenvalues is then taken as the value of Max(Im[$E$]). Below are the details of the tight-binding model matrix construction as well as the subsequent numerical diagonalization.

The illustrated model in our specified model is a coupled bilayer system where each layer is given by the well-known Qi-Wu-Zhang Chern model, whose Bloch Hamiltonian is written as 
\begin{equation}
        H_\text{Ch}(\boldsymbol{k},\gamma) = (m +  \cos k_x +  \cos k_y) \sigma_x + (\mathrm{i} \gamma + \sin k_y) \sigma_y + ( \sin k_x) \sigma_z.
    \label{req:onelayer-bloch}
\end{equation}

After imposing open boundary conditions (OBC) in both $x$ and $y$ directions, we can obtain the double-OBC Hamiltonian representing a single rectangular lattice with sizes $L_x$ and $L_y$, which is given by
\begin{equation}
    \begin{array}{r}
        H_\text{Ch}(\gamma)=\sum_x \sum_y\left\{\left[\hat{c}_{x, y+1}^{\dagger} \frac{\left(\sigma_x+\mathrm{i} \sigma_z\right)}{2} \hat{c}_{x, y}+\text { H.c. }\right]\right. \\
        +\left[\hat{c}_{x+1, y}^{\dagger} \frac{\left(\sigma_x+\mathrm{i} \sigma_y\right)}{2} \hat{c}_{x, y}+\text { H.c. }\right] \\
        \left.+\hat{c}_{x, y}^{\dagger}\left(m \sigma_x+\mathrm{i} \gamma \sigma_y\right) \hat{c}_{x, y}\right\},
        \end{array}
\end{equation}
where $\hat{c}_{x, y}\left(\hat{c}_{x, y}^{\dagger}\right)$ annihilates (creates) a fermion with two internal degrees of freedom on site $(x, y)$. The $\sigma$'s are the Pauli matrices acting on the internal degrees of freedom. The shape of $H_\text{Ch}(\gamma)$ is $2L_x L_y$ (the number 2 is from the size of the Pauli matrices). In the matrix representation, taking $L_x=3, L_y=2$ as an example, labelling the unit cells in indices 1, 2, 3 (on the first row), and 4, 5, 6 (on the second row). In this case, this double OBC Hamiltonian is given by
\begin{equation}
    H_\text{Ch}(\gamma)=\left(\begin{array}{cccccc}
        m \sigma_x+\mathrm{i} \gamma \sigma_y & \frac{\left(\sigma_x+\mathrm{i} \sigma_y\right)}{2} & 0 & \frac{\left(\sigma_x+\mathrm{i} \sigma_z\right)}{2} & 0 & 0 \\
        \text { H.c. } & m \sigma_x+\mathrm{i} \gamma \sigma_y & \frac{\left(\sigma_x+\mathrm{i} \sigma_y\right)}{2} & 0 & \frac{\left(\sigma_x+\mathrm{i} \sigma_z\right)}{2} & 0 \\
        0 & \text { H.c. } & m \sigma_x+\mathrm{i} \gamma \sigma_y & 0 & 0 & \frac{\left(\sigma_x+\mathrm{i} \sigma_z\right)}{2} \\
        \text { H.c. } & 0 & 0 & m \sigma_x+\mathrm{i} \gamma \sigma_y & \frac{\left(\sigma_x+\mathrm{i} \sigma_y\right)}{2} & 0 \\
        0 & \text { H.c. } & 0 & \text { H.c. } & m \sigma_x+\mathrm{i} \gamma \sigma_y & \frac{\left(\sigma_x+\mathrm{i} \sigma_y\right)}{2} \\
        0 & 0 & \text { H.c. } & 0 & \text { H.c. } & m \sigma_x+\mathrm{i} \gamma \sigma_y
        \end{array}\right)
\end{equation}

Thus, the bilayer system is then given by
\begin{equation}
    \begin{aligned}
        \hat{H}_{\text{two-layer}}(\gamma,\mu_0) &= \begin{pmatrix}
            \hat{H}_\text{Ch}(\gamma) & \mu_0\,\mathbb{I} \\
            \mu_0\,\mathbb{I} & \hat{H}_\text{Ch}(-\gamma)
        \end{pmatrix}.
    \end{aligned}
\end{equation}
By numerically diagonalizing the Hamiltonian $\hat{H}_{\text{two-layer}}(\gamma,\mu_0)$, we can obtain the eigenvalues $E$. The maximum imaginary part of the eigenvalues for topological modes is then taken as the value of Max(Im[$E$]) in the topological band. 

\color{black}

\section{Estimation of the effective parameters characterizing the dynamically generated edge states}

\subsection{Estimation of the decay length for dynamically generated edge states}

In our topologically guided gain mechanism, the inverse state decay length (inverse skin depth) $\kappa$ is the key measure of the asymmetric directionality of the non-Hermitian couplings. We first describe it for our bilayer non-Hermitian Chern model $\bar H_\text{two-layer}$ and then show how we can estimate the effective $\kappa$ for the dynamically evolved state, which is a superposition of many such modes.

Since the asymmetry is in the $x$-direction, we consider $x$-open boundary conditions (OBCs), such that we obtain effective 1D chains with energy dispersion
\begin{equation}
  E^2(z,k_y,\gamma)=\left(m+\cos k_y+\gamma+z^{-1}\right)\left(m+\cos k_y-\gamma+z\right),
\end{equation}
where $z=e^{-\kappa}e^{ik_x}$. We rewrite this as 
\begin{equation}
  (m+\cos k_y+\gamma)z^2+\left((m+\cos k_y)^2-\gamma^2+1+\sin k_y^2-E^2\right)z+(m+\cos k_y-\gamma)=0.
\end{equation}
such that solutions $z=z_{1,2}$ satisfying $|z_1|=|z_2|$ are constrained~\cite{lee2019anatomy,yao2018edge,yokomizo2019non,yang2020non,lee2020unraveling} to the OBC energies $E$ and satisfy 
\begin{equation}
  z=-\sqrt{\frac{m+\cos k_y-\gamma}{m+\cos k_y+\gamma}},
\end{equation}
such that $\kappa=-\ln(|z|)$ is given by
\begin{equation}
  \kappa=\ln\left(\left|\sqrt{\frac{m+\cos k_y+\gamma}{m+\cos k_y-\gamma}}\right|\right).
\end{equation}

However, in the system where we dynamically excite many modes with different real and imaginary parts of eigenenergies which may correspond to different decay lengths, the decay length is not well defined. In this case, we can define the decay length $\kappa$ as the average of the decay length of all the excited modes, corresponding to an effective $\kappa$ and corresponding effective $k_y$. 

\begin{figure}[!htp]
    \centering
    \includegraphics[width=0.4\textwidth]{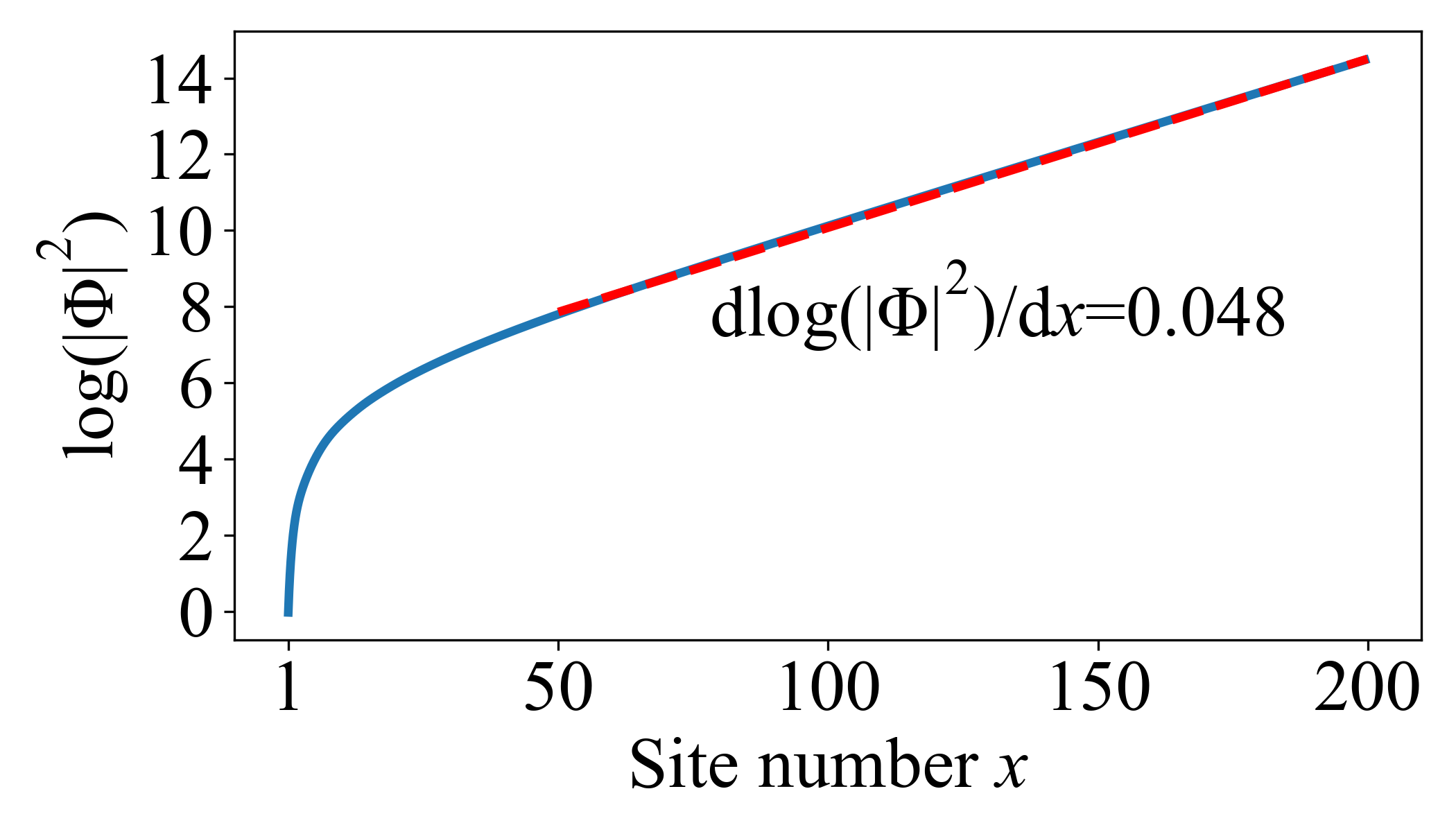}
    \caption{The blue curve is the logarithmic amplitude squared of the state $\Phi(\bold r,t)$ excited by the source $\eta(\mathbf r,t)$. The red dashed line is a linear fit with a slope of 0.048, which should correspond to $2\kappa$. This yields an effective averaged $\kappa=0.024$. This simulation is conducted in the rectangular lattice with $L_x=200, L_y=5, \gamma=0.02$. 
		}
    \label{smfig:kappa}
\end{figure}

To empirically obtain the spatial decay rate $\kappa$ of the evolved state $|\Phi(\bold r, t)\rangle$ which was excited by the source $\ket{\eta(\mathbf r,t)}$, we plot the logarithmic amplitude squared of $\Phi(\bold r, t)$ versus site number, as shown in Fig.~\ref{smfig:kappa}. The red dashed line is plotted with the slope being 0.048, which should correspond to $2\kappa$. As such, we obtain the effective averaged $\kappa=0.024$ which corresponds to an effective $k_y\approx 0.56\pi$.

\subsection{Estimation of the wavepacket velocity for dynamically generated edge states}
\begin{figure}[!htp]
  \centering
  \includegraphics[width=0.6\textwidth]{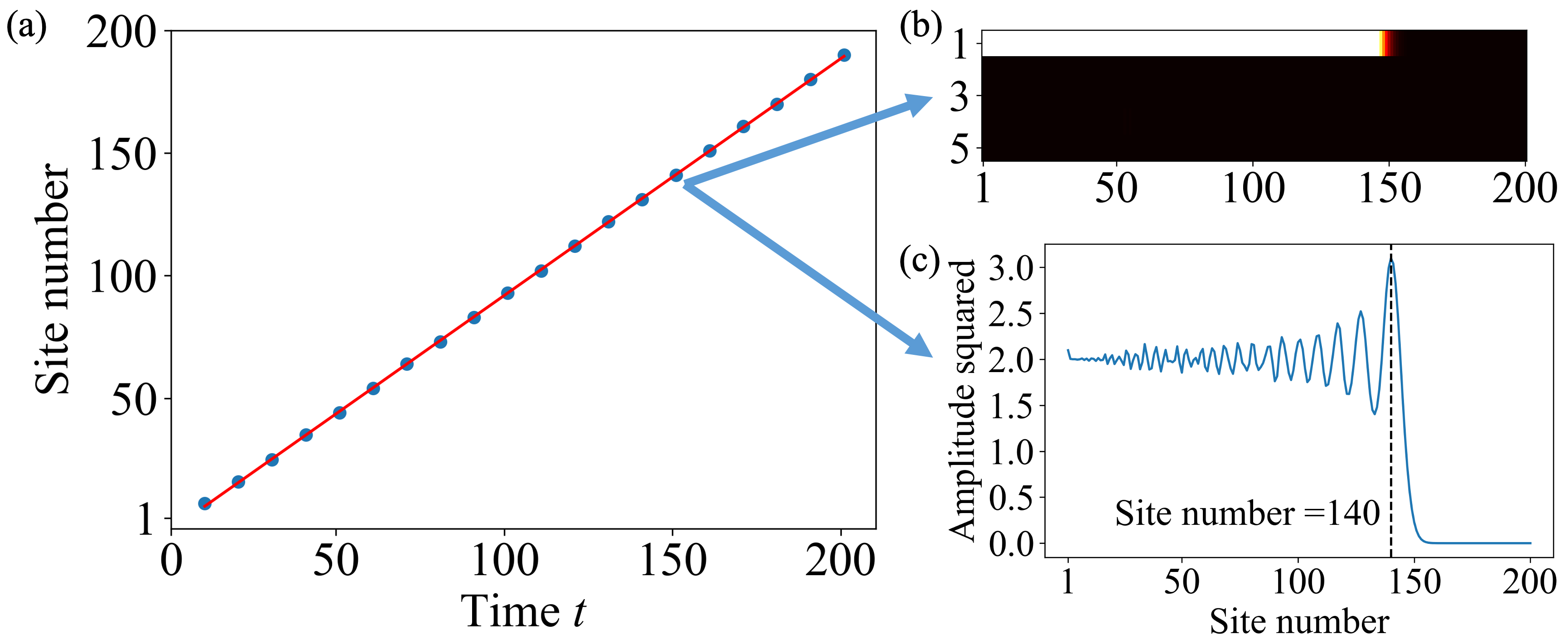}
  \caption{Here, we have set the rectangular lattices with $L_x=200, L_y=5, \gamma=0$. (a) The blue dots are positions where the wavepacket resides when time slots are $t=10, 20, ..., 200$. And the slope of the linear fit (red) is 0.96, which is the velocity of the wavepacket. (b) and (c) are showing how the site number for a certain time slot $t=150$ is determined. (b) Amplitude squared of the state at time $t=150$. (c) The site number for time $t=150$ is determined by the maximum of the amplitude squared. The dashed line indicates that the site number is 140. }
  \label{smfig:velocity}
\end{figure}

The rate of topologically guided gain is proportional to the velocity of the chirally propagating edge states, which we can measure through the advancement of the dynamical wavefront generated in the simulation. Since the velocity of the chiral modes is not affected by the non-Hermiticity or the width of the system, at least for our $\hat H_\text{two-layer}$ model, we can estimate the wavepacket velocity using the Hermitian case with $L_x=120$, where the state suffers no attenuation and is hence more easily measured, as shown in the following Fig.~\ref{smfig:velocity}. By plotting the amplitude squared of state at each time slice $t=10, 20, ..., 200$, we can obtain the movement of the wavepacket (Fig.~\ref{smfig:velocity}(a)). Figs.~\ref{smfig:velocity}(b) and (c) show the amplitude profile of an illustrative snapshot. The slope, 0.96, fitted shown in the red line in Fig.~\ref{smfig:velocity}(a) indicates the velocity of the chiral propagation of the wavepacket. 

\color{black}
\subsection{The effects of the effective coupling $\mu$ when $\mu\geq1$ and $\mu\to 0$}

In Eq. (1) of the main text, we have claimed that the critical width of our percolated system is 
\begin{equation}
  x^*=vt^*\approx -\kappa^{-1}\log\mu.
  \label{smeq:xcrit}
  \end{equation}
$v$ is the velocity of the topological edge state, $t^*$ is the time at which the topologically guided gain is triggered, and $\kappa$ is the inverse decay length of the edge state, which we can obtain from the previous subsections in the Supplemental Material. 

We note that the threshold $x^*$ is estimated as $-\text{log}\mu$, and this estimation becomes negative when $\mu \geq 1$. This leads to the conclusion that the PT breaking can not happen for any small sizes. However, if the interlayer coupling $\mu$ is sufficiently large, the opposite hopping asymmetry channels in two layers will cancel each other's skin effect, thereby hindering the manifestation of the topologically guided gain. Consequently, in such scenarios, PT symmetry breaking fails to occur due to the suppression of the topologically guided gain. In such a case, we still necessitate the interlayer coupling $\mu$ to remain sufficiently small as well as sufficiently large to avoid cancelling the skin effect, yet it must also be capable of inducing PT symmetry breaking through the topologically guided gain. In this context, percolation remains relevant.

We also note that on the opposite limit that $\mu\to 0$, $x^* \to\infty$. This means in the thermodynamic limit, only infinitesimal non-Hermiticity is needed to trigger PT breaking, which is consistent with the conclusion of the Ref.~\cite{song2022non}. However, the mechanism driving PT symmetry breaking differs significantly between our work and that of Ref.~\cite{song2022non}. In our research, PT symmetry breaking is instigated by the novel concept of topologically guided gain, a mechanism previously unexplored in dynamic contexts and consistently supported by exact diagonalization results. The agreement between our dynamic simulation findings and the theoretical predictions derived from numerical diagonalization of Hamiltonians reinforces this novel mechanism. In contrast, Ref.~\cite{song2022non} focus on non-Bloch PT symmetry within two and higher dimensions, revealing notably distinct behaviors compared to its one-dimensional counterpart, which is a dimensional surprise.

Their model comprises a single-layer 2D non-Hermitian lattice, which is 
\begin{equation}
    \begin{array}{r}
        H(\mathbf{k})=(t-\gamma) e^{i k_x}+(t+\gamma) e^{-i k_x}+(t-\gamma) e^{i k_y}+ \\
        (t+\gamma) e^{-i k_y}+s\left(e^{i k_x}+e^{-i k_x}\right)\left(e^{i k_y}+e^{-i k_y}\right). 
    \end{array}
\end{equation}
This model showcases complexity due to the coupling $s$, which interconnects multiple competitive skin effect channels, leading to dimensional surprise. However, the illustrated Hamiltonian in our model represents a simpler bilayer system. Here, the interlayer coupling $\mu$ serves as the sole parameter connecting the two layers, each hosting skin effect channels either on the left or right (no multiple competitive skin effect channels as those in Song et al.), as one can see in the following equation:
\begin{equation}
    \begin{aligned}
        \hat{H}_{\text{two-layer}}(\boldsymbol{k},\gamma,\mu) &= \begin{pmatrix}
            \hat{H}_\text{Ch}(\boldsymbol{k},\gamma) & \mu_0\,\mathbb{I} \\
            \mu_0\,\mathbb{I} & \hat{H}_\text{Ch}(\boldsymbol{k},-\gamma)
        \end{pmatrix},\\
        H_\text{Ch}(\boldsymbol{k},\gamma) &= (m +  \cos k_x +  \cos k_y) \sigma_x \\
        &\quad + (\mathrm{i} \gamma + \sin k_y) \sigma_y + ( \sin k_x) \sigma_z,
    \end{aligned}
\end{equation}

\section{Effective theory in the subspace spanned only by topological edge states}

In the main text, we have shown that the topological edge states are the main contributors to the topologically guided gain. Here, we provide a detailed analysis of the effective Hamiltonian in the basis of the topological edge modes. The effective Hamiltonian can be reduced to 
\begin{equation}
    \hat{H}_{\text{eff}} = \begin{pmatrix}
        vk-i\kappa & \mu_0\ \\
        \mu_0 & vk+i\kappa
    \end{pmatrix}.
    \label{req:eff}
\end{equation}
$vk$ denotes the topological edge modes traveling in a clockwise direction, while $\pm i\kappa$ leads to the growth or decay of the states. The interlayer coupling, represented by $\mu_0$, serves as a parameter linking the two layers. If we throw an initial state $\psi(t=0)$ on the upper layer, which is written as $(1,0)^T$ in this basis, we can obtain the state evolutions at any time $t$ by solving $(\psi^U,\psi^L)^T=\exp(-i\hat{H}_{\text{eff}}t)(1,0)^T$, which are
\begin{equation}
    \psi^U(t)=\frac{{e^{-t (\sqrt{\kappa^2 - \mu_0^2} + i vk)} \left(\kappa - e^{2 \sqrt{\kappa^2 - \mu_0^2} t} \kappa + \left(1 + e^{2 \sqrt{\kappa^2 - \mu_0^2} t}\right) \sqrt{\kappa^2 - \mu_0^2}\right)}}{{2 \sqrt{\kappa^2 - \mu_0^2}}},\quad \mu_0<\kappa\label{eq:eff_upper}
\end{equation}
and 
\begin{equation}
    \psi^L(t)=-\frac{{i e^{-t (\sqrt{\kappa^2 - \mu_0^2} + i vk)} \left(-1 + e^{2 \sqrt{\kappa^2 - \mu_0^2} t}\right) \mu_0}}{{2 \sqrt{\kappa^2 - \mu_0^2}}},\quad \mu_0<\kappa.\label{eq:eff_lower}
\end{equation}
Here, we have assumed that $\mu_0<\kappa$, which corresponds to the case that the interlayer couping $\mu_0$ is small enough.

\begin{figure}[!htp]
  \centering
  \subfigure[$\mu_0<\kappa$, absolute amplitude of Eqs.~\ref{eq:eff_upper} and \ref{eq:eff_lower}. ]{\includegraphics[width=0.48\textwidth]{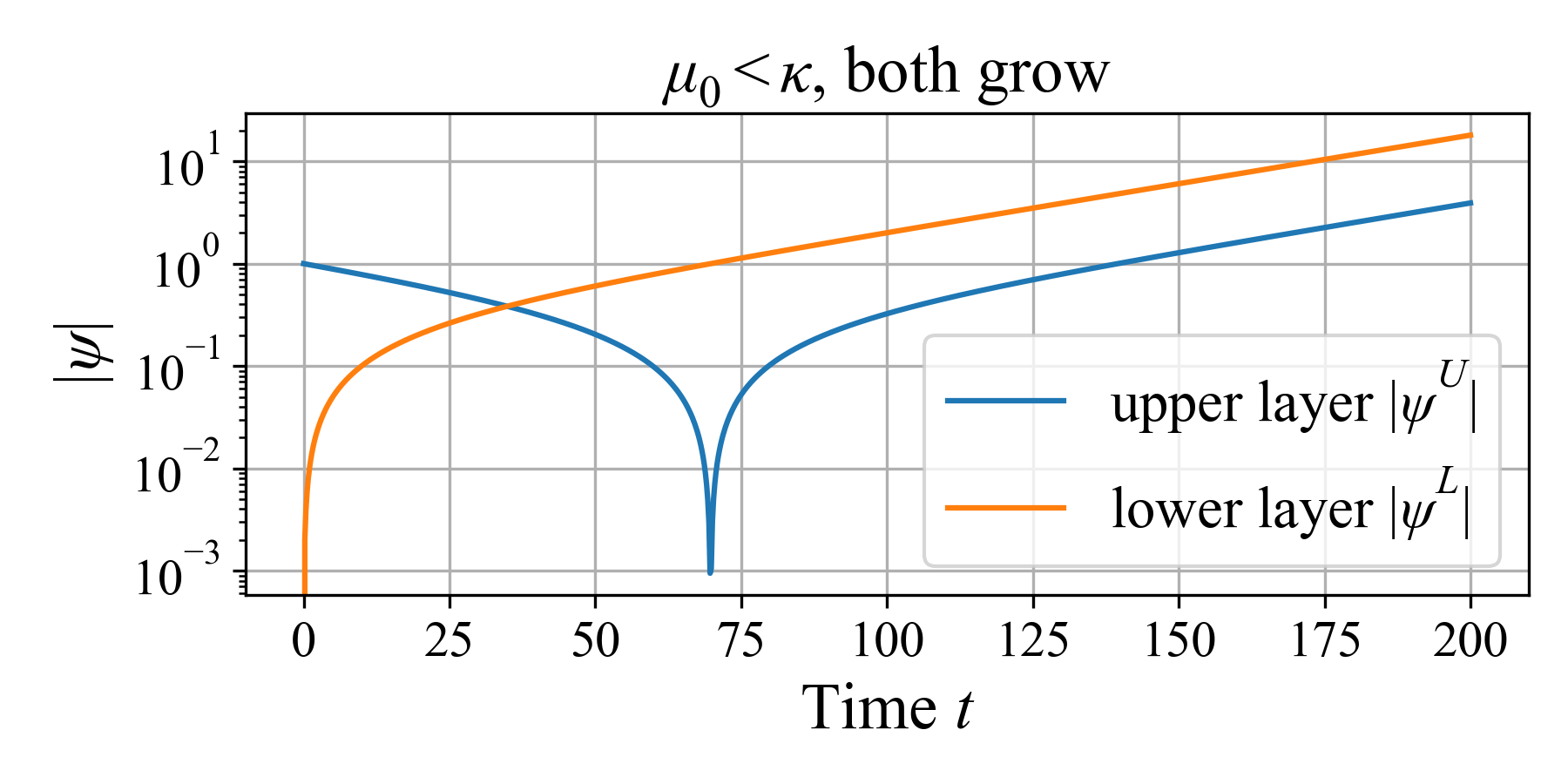}}
  \subfigure[$\mu_0>\kappa$, absolute amplitude of Eqs.~\ref{eq:eff_upper_large_mu} and \ref{eq:eff_lower_large_mu}. ]{\includegraphics[width=0.48\textwidth]{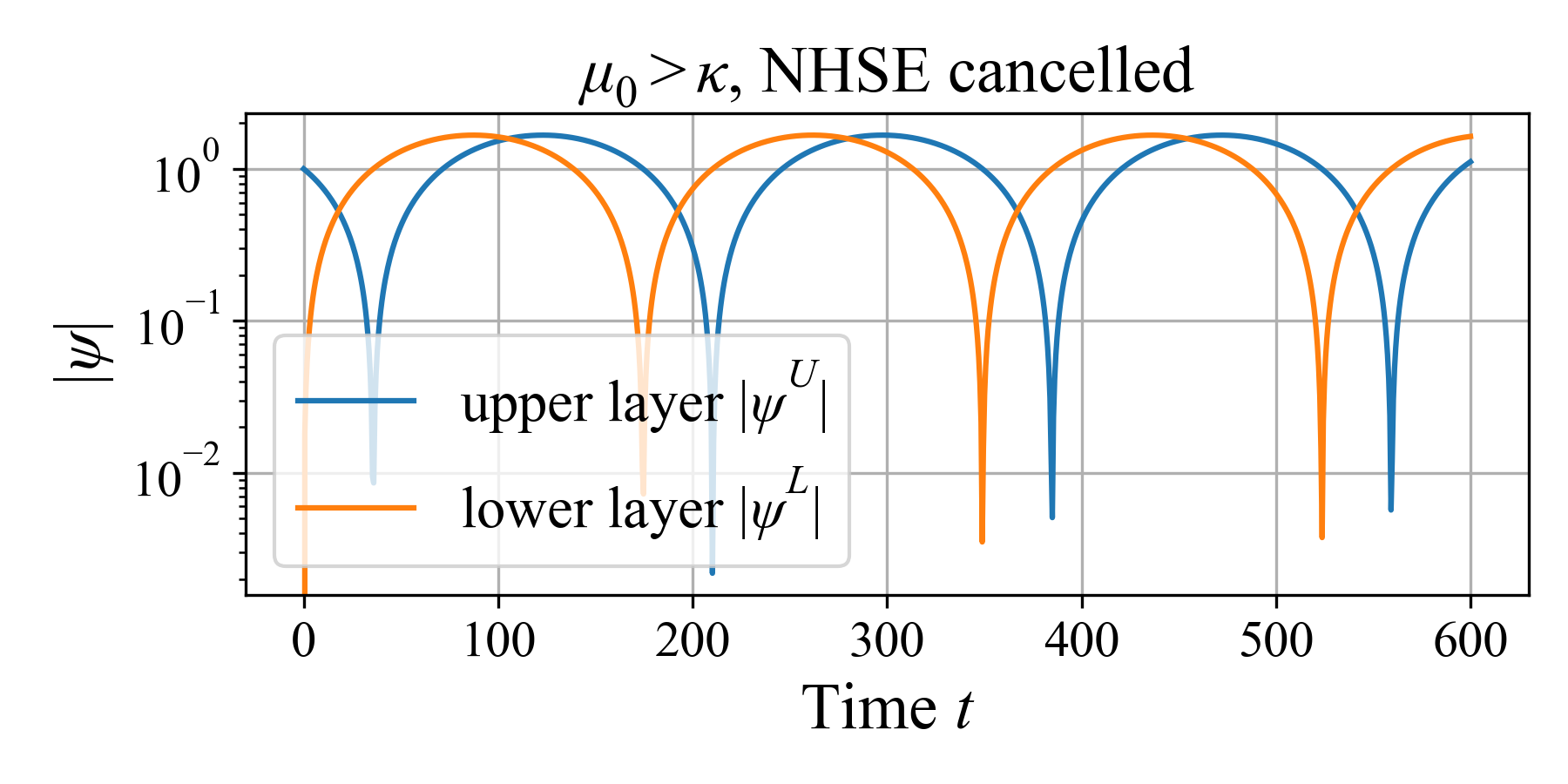}}
  \caption{\textcolor{black}{A logarithmic graph depicts the absolute amplitudes of the wavepacket on two layers: the upper layer $|\psi^U|$ (shown in blue) and the lower layer $|\psi^L|$ (depicted in yellow) in (a) Eqs.~\ref{eq:eff_upper} and \ref{eq:eff_lower}, and (b) Eqs.~\ref{eq:eff_upper} and \ref{eq:eff_lower}. Initially, the wavepacket $(1,0)^T$ is solely on the upper layer. (a) The parameters are $\kappa=0.024,\mu_0=0.01$. Over time, the wavepacket on the upper layer shrinks initially due to the NHSE's opposing direction. However, as the wavepacket on the lower layer expands, their combined effect becomes apparent. Ultimately, both wavepackets on the upper and lower layers grow in magnitude. (b) The parameters are $\kappa=0.024,\mu_0=0.03$. Over time, we found wavepackets on both layers oscillate, without any amplification or attenuation with $\mu_0>\kappa$, which is very different from the result in Fig.~\ref{sfig:eff_topo}(a) with $\mu_0<\kappa$. In both subfigures, the absolute value of $vk$ does not matter in this context. } }
  \label{sfig:eff_topo}
\end{figure}

Plotting the absolute amplitudes of the wavepacket on both layers in Eqs.~\ref{eq:eff_upper} and \ref{eq:eff_lower} allows us to track its evolution, where we have used the same parameters as in Fig.~2 of the main text: $\kappa=0.024,\mu_0=0.01$ (here the absolute value of $vk$ will not affect $|\psi^U|$ or $|\psi^L|$). As depicted in Fig.~\ref{sfig:eff_topo}(a), initially, the wavepacket on the upper layer contracts due to the opposing direction of the NHSE. However, with the expansion of the wavepacket on the lower layer, their mutual influence becomes evident. Ultimately, both wavepackets on the upper and lower layers increase in magnitude. The transition width $x^*\sim70$ is very close to the one in Fig.~2 ($x^*\sim 60$) of the main text, as we still can not ignore the effect of the bulk states, which is the reason why the transition width is not exactly the same.

On the other hand, let us discuss the case where $\mu_0>\kappa$. In this case, the state evolutions are given by
\begin{equation}
  \psi^U(t)=\frac{{e^{-t (i\sqrt{\mu_0^2-\kappa^2} + i vk)} \left(\kappa - e^{2i \sqrt{\mu_0^2-\kappa^2} t} \kappa + \left(1 + e^{2 i\sqrt{\mu_0^2-\kappa^2} t}\right) i\sqrt{\mu_0^2-\kappa^2}\right)}}{{2i \sqrt{\mu_0^2-\kappa^2}}},\quad \mu_0>\kappa\label{eq:eff_upper_large_mu}
\end{equation}
and
\begin{equation}
  \psi^L(t)=-\frac{{i e^{-t (i\sqrt{\mu_0^2-\kappa^2} + i vk)} \left(-1 + e^{2i \sqrt{\mu_0^2-\kappa^2} t}\right) \mu_0}}{{2i \sqrt{\mu_0^2-\kappa^2}}},\quad \mu_0>\kappa.\label{eq:eff_lower_large_mu}
\end{equation}


Similarly, we plot the absolute amplitudes of the wavepacket on both layers in Eqs.~\ref{eq:eff_upper_large_mu} and \ref{eq:eff_lower_large_mu} to track its evolution, where we have used the same parameters as in Fig.~2 of the main text: $\kappa=0.024$, but now we choose $\mu_0=0.03$ to show NHSE cancellation. As depicted in Fig.~\ref{sfig:eff_topo}(b), wavepackets on both layers oscillate, without any amplification or attenuation, which is very different from the result in Fig.~\ref{sfig:eff_topo} where $\mu_0<\kappa$. This is consistent with the fact that the NHSE is cancelled when $\mu_0>\kappa$.

\textcolor{black}{It is noted that the PT symmetry breaking does not happen in a short period of time. It happens after evolving a certain time, which requires a sufficiently long length in one direction. Therefore, the PT symmetry breaking is indeed associated with length, which agrees with the numerical results in the main text.}

\section{Effect of interlayer coupling nonlinearity on topologically guided gain}

In the main text, we have assumed that the interlayer coupling is linear for both layers. We would like firstly clarift that this linearity comes from the linearly in the coupling, which leads to the linear differential equation shown both in the main text and the Supplemental Material. 

However, even if we have a nonlinear coupling, the topologically guided gain effect will still hold, for example, now we have this nonlinear differential equation when $\delta\neq0$
\begin{equation}
    i\frac{d}{dt}\begin{pmatrix}
        \psi^U(t) \\
        \psi^L(t)
    \end{pmatrix}=
    \begin{pmatrix}
        \hat{H}^U & \mu_0\,\mathbb{I} \\
        \mu_0\,\mathbb{I} & \hat{H}^L
    \end{pmatrix}
    \begin{pmatrix}
        \left(\psi^U(t)\right)^{(1+\delta)} \\
        \left(\psi^L(t)\right)^{(1+\delta)}
    \end{pmatrix}
\end{equation}

For simplicity, we study this nonlinear differential equation in the spanned topological states subspace, which can be given by 
\begin{equation}
    i\frac{d}{dt}\begin{pmatrix}
        \psi^U(t) \\
        \psi^L(t)
    \end{pmatrix}=
    \begin{pmatrix}
        vk-i\kappa & \mu_0 \\
        \mu_0 & vk+i\kappa
    \end{pmatrix}
    \begin{pmatrix}
        \left(\psi^U(t)\right)^{(1+\delta)} \\
        \left(\psi^L(t)\right)^{(1+\delta)}
    \end{pmatrix}
\end{equation}

\begin{figure}[!htp]
    \centering
    \subfigure[$|\psi^U(t)|$]{\includegraphics[width=0.48\textwidth]{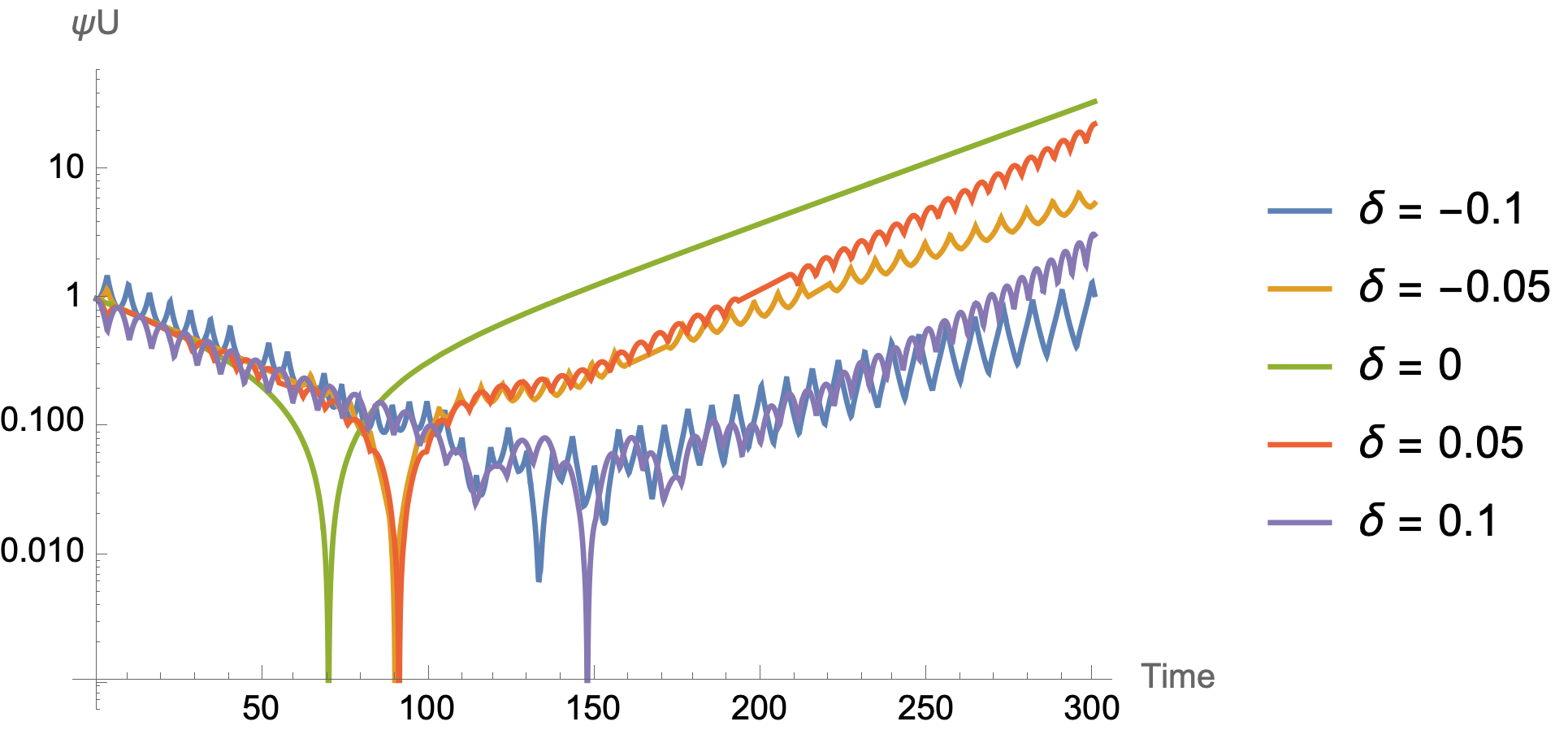}}
    \subfigure[$|\psi^L(t)|$]{\includegraphics[width=0.48\textwidth]{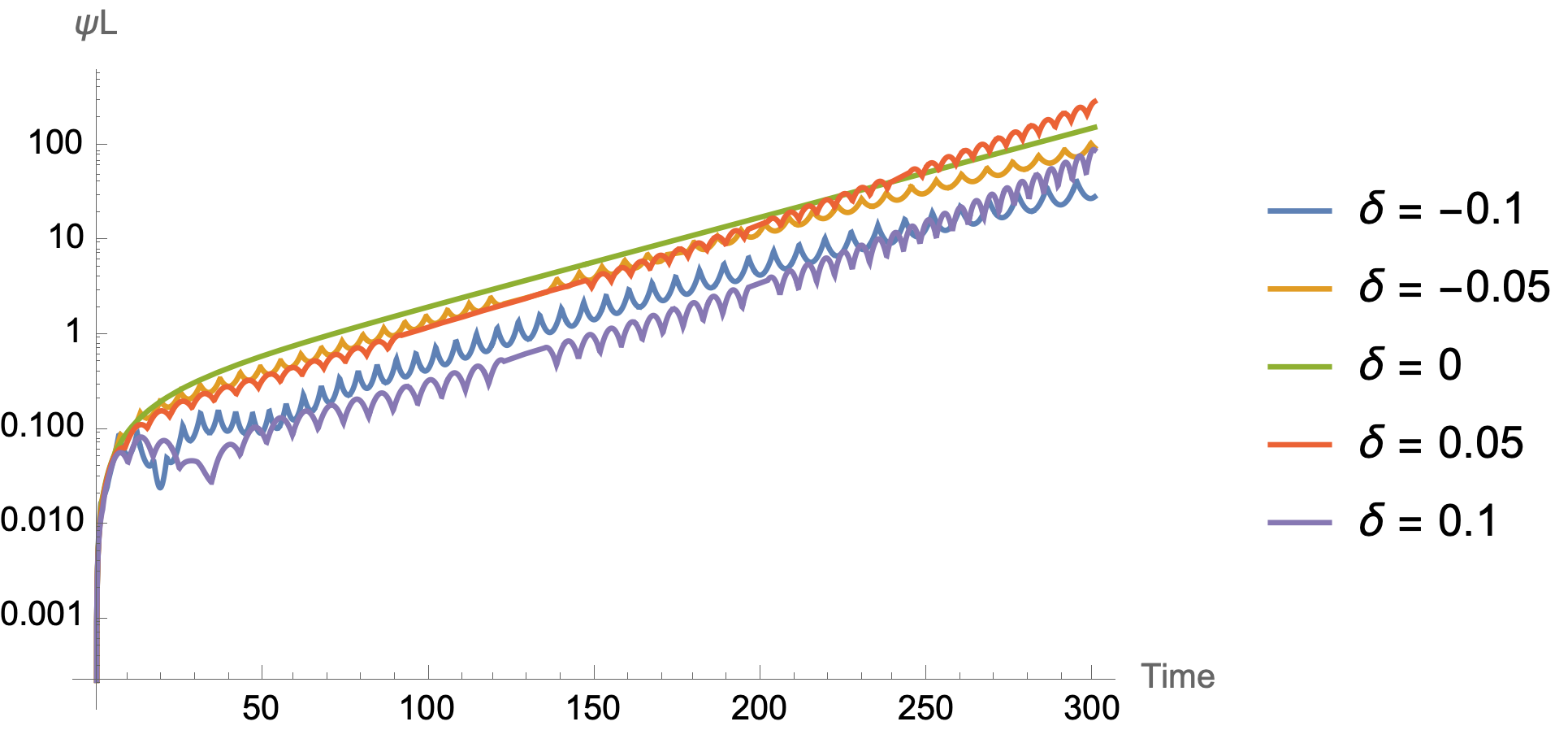}}
    \caption{\textcolor{black}{A logarithmic graph depicts the absolute amplitudes of the wavepacket on two layers against different nonlinearity $\delta=-0.1, -0.05, 0, 0.05, 0.1$ in various colors. (a) the upper layer $|\psi^U|$, and (b) the lower layer $|\psi^L|$. Initially, the wavepacket $(1,0)^T$ is solely on the upper layer. We found that, over time, we still observe topologically guided gain on the upper layer. Further, the nonlinearity mainly leads to the shift of the transition point. The parameters are $\kappa=0.024,\mu_0=0.01, vk=1$. } }
    \label{sfig:nonlinear}
  \end{figure}

Here, we solve the differential equations and provide the numerical results of the absolute values for the wavepacket amplitude on the upper layer and the lower layer, i.e., $\psi^U(t)$ and $\psi^L(t)$ in Fig.~\ref{sfig:nonlinear}, where we have set the parameters as $\kappa=0.024,\mu_0=0.01,\delta=-0.1, -0.05, 0, 0.05, 0.1$. We can see that the topologically guided gain effect still holds, even if we have a nonlinear coupling which caused the shift on the transition point where we can start to see the topologically guided gain.

\color{black}
\section{Generation of the disordered islands through a spatially correlated random landscape}

The disordered island configurations in our work are generated by filling a random landscape $m(\bold r)$ up to a ``sea level'' of $m^*$, such that regions with $m(\bold r)\geq m^*$ are dry and designed as the islands. In the following, we describe how we generate the random landscape such that the island sizes and the smoothness of their boundaries are approximately tunable via $m^*$.

\begin{figure}[!htp]
  \centering
  \subfigure[]{\includegraphics[width=0.21\textwidth]{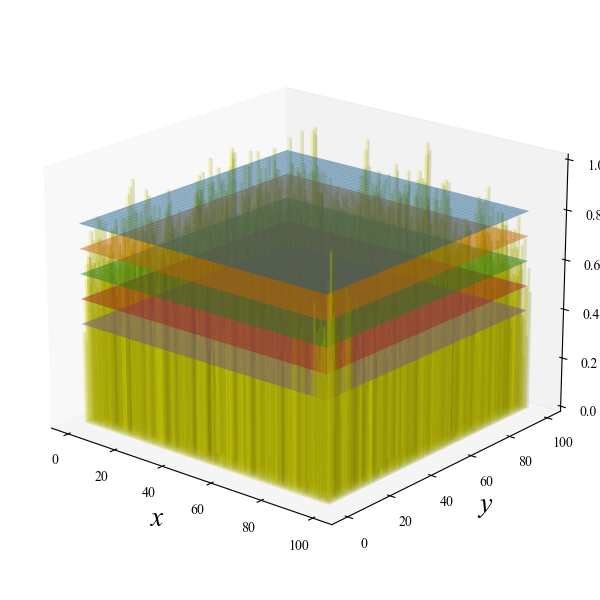}}
  \subfigure[]{\includegraphics[width=0.15\textwidth]{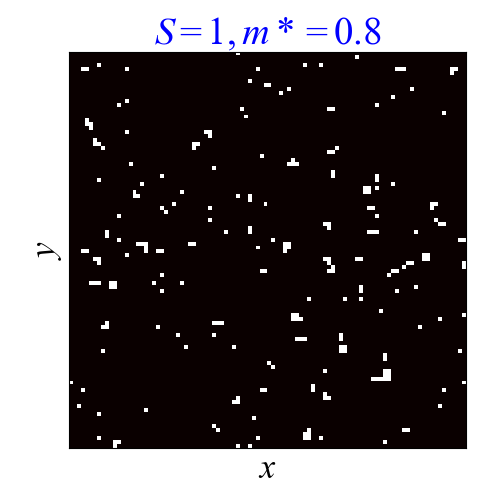}}
  \subfigure[]{\includegraphics[width=0.15\textwidth]{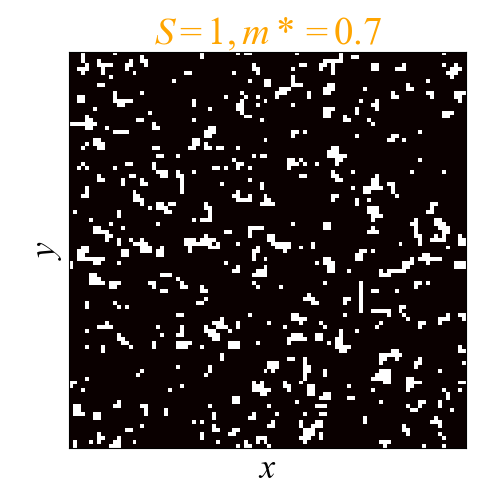}}
  \subfigure[]{\includegraphics[width=0.15\textwidth]{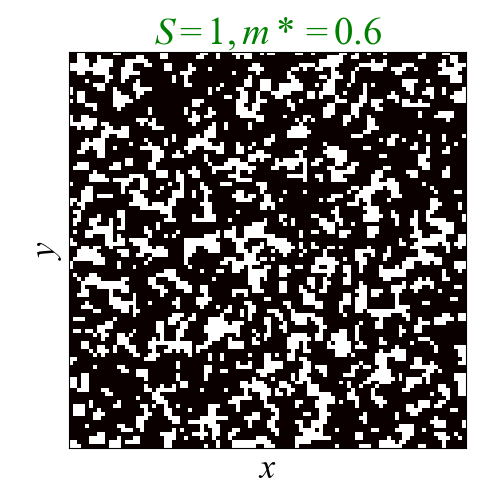}}
  \subfigure[]{\includegraphics[width=0.15\textwidth]{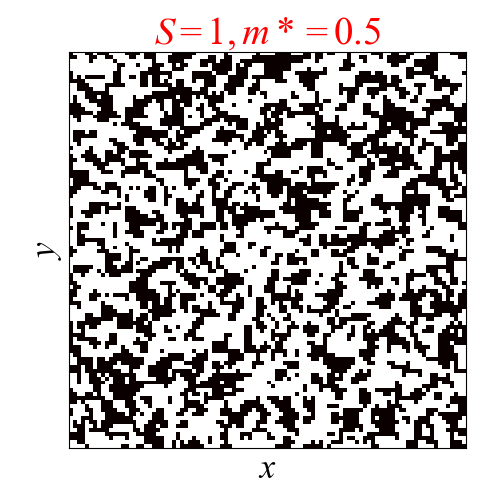}}
  \subfigure[]{\includegraphics[width=0.15\textwidth]{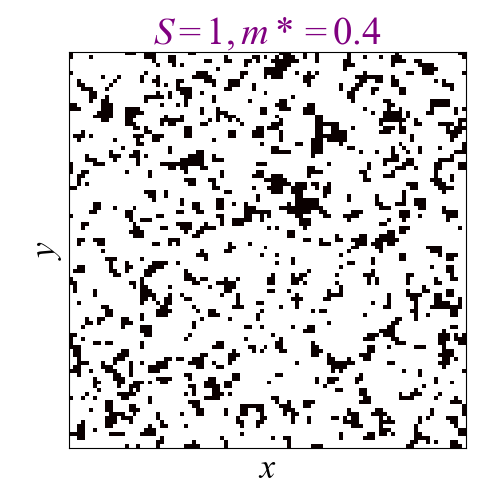}}
  \caption{(a) 3D landscape with smoothness  $S=1$, and $L_x=L_y=100$. (b)--(f) The disordered texture defined by the intersection of the landscape and the ``sea levels'' (b) $m^*=0.8$, (c) $m^*=0.7$, (d) $m^*=0.6$, (e) $m^*=0.5$, and (f) $m^*=0.4$. }
  \label{smfig:S=1}
\end{figure}

\begin{figure}[!htp]
  \centering
  \subfigure[]{\includegraphics[width=0.21\textwidth]{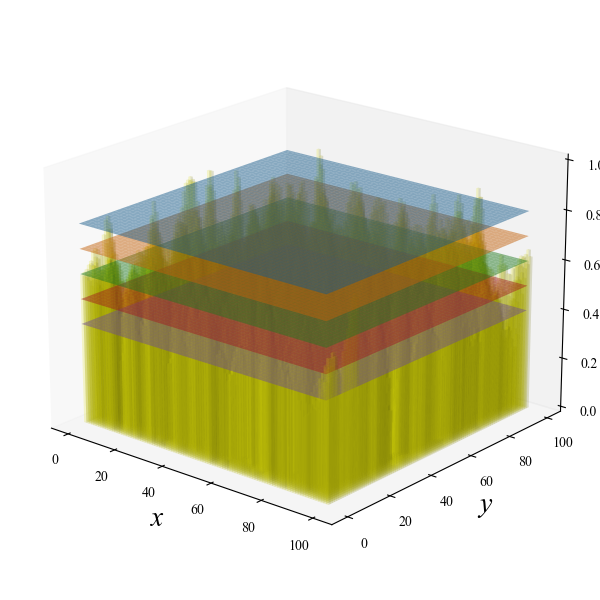}}
  \subfigure[]{\includegraphics[width=0.15\textwidth]{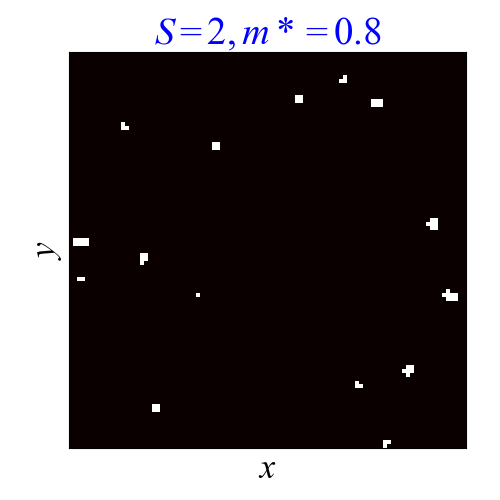}}
  \subfigure[]{\includegraphics[width=0.15\textwidth]{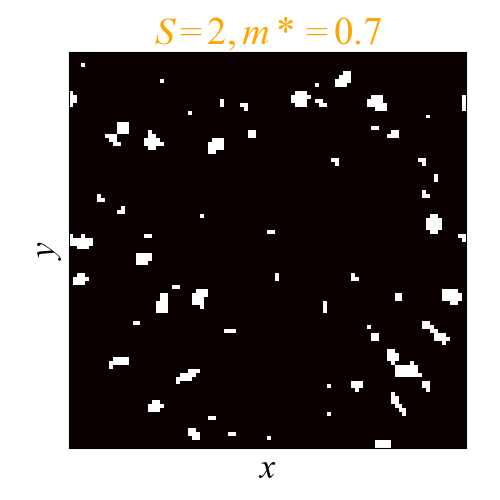}}
  \subfigure[]{\includegraphics[width=0.15\textwidth]{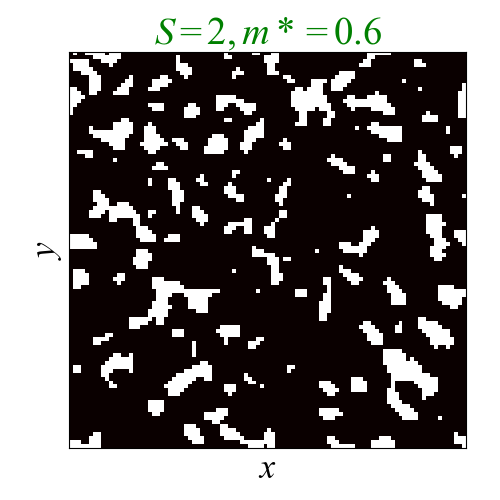}}
  \subfigure[]{\includegraphics[width=0.15\textwidth]{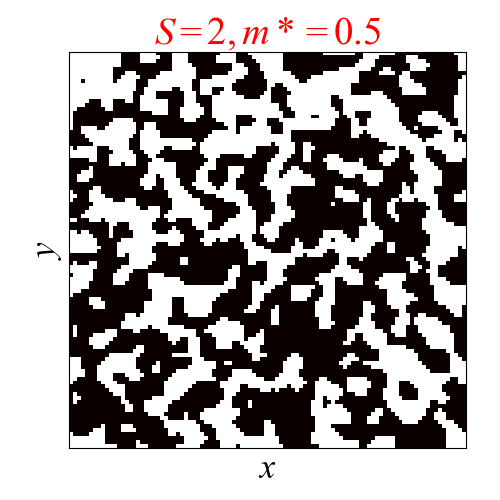}}
  \subfigure[]{\includegraphics[width=0.15\textwidth]{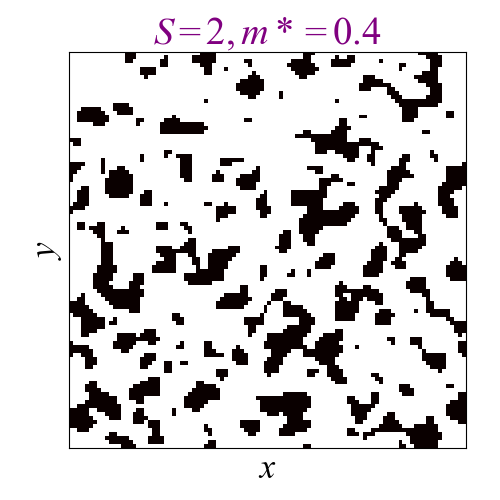}}
  \caption{(a) 3D landscape with smoothness  $S=2$, and $L_x=L_y=100$. (b)--(f) The disordered texture defined by the intersection of the landscape and the ``sea levels'' (b) $m^*=0.8$, (c) $m^*=0.7$, (d) $m^*=0.6$, (e) $m^*=0.5$, and (f) $m^*=0.4$. }
  \label{smfig:S=2}
\end{figure}

\begin{figure}[!htp]
  \centering
  \subfigure[]{\includegraphics[width=0.21\textwidth]{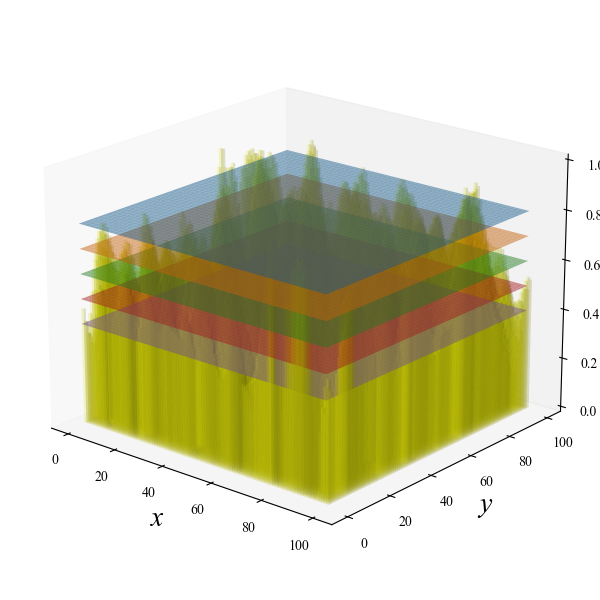}}
  \subfigure[]{\includegraphics[width=0.15\textwidth]{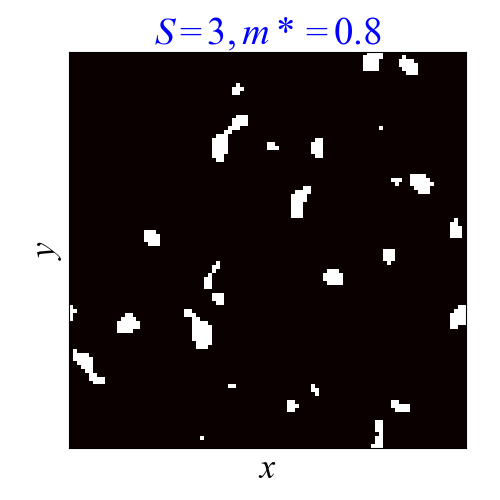}}
  \subfigure[]{\includegraphics[width=0.15\textwidth]{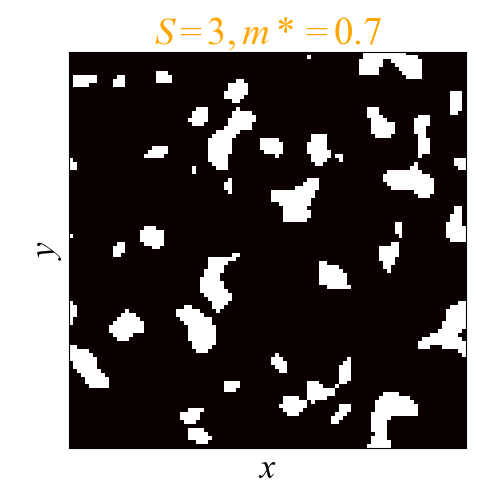}}
  \subfigure[]{\includegraphics[width=0.15\textwidth]{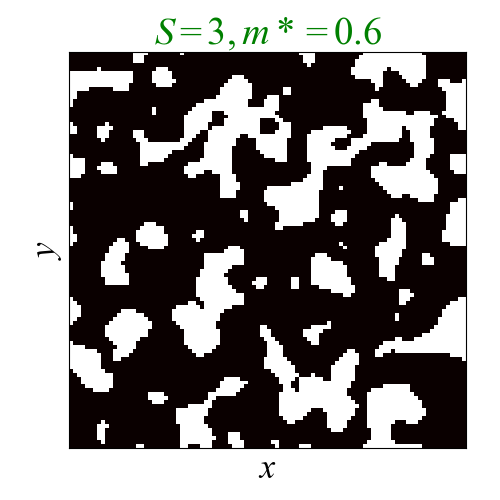}}
  \subfigure[]{\includegraphics[width=0.15\textwidth]{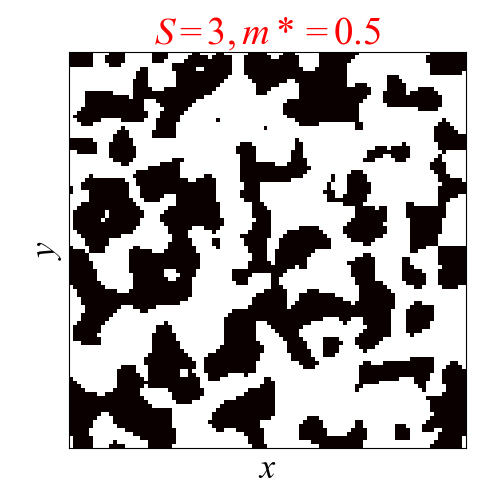}}
  \subfigure[]{\includegraphics[width=0.15\textwidth]{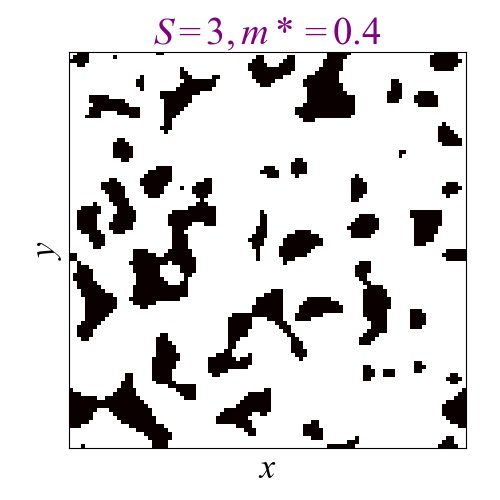}}
  \caption{(a) 3D landscape with smoothness  $S=3$, and $L_x=L_y=100$. (b)--(f) The disordered texture defined by the intersection of the landscape and the ``sea levels'' (b) $m^*=0.8$, (c) $m^*=0.7$, (d) $m^*=0.6$, (e) $m^*=0.5$, and (f) $m^*=0.4$. }
  \label{smfig:S=3}
\end{figure}

\begin{figure}[!htp]
  \centering
  \subfigure[]{\includegraphics[width=0.21\textwidth]{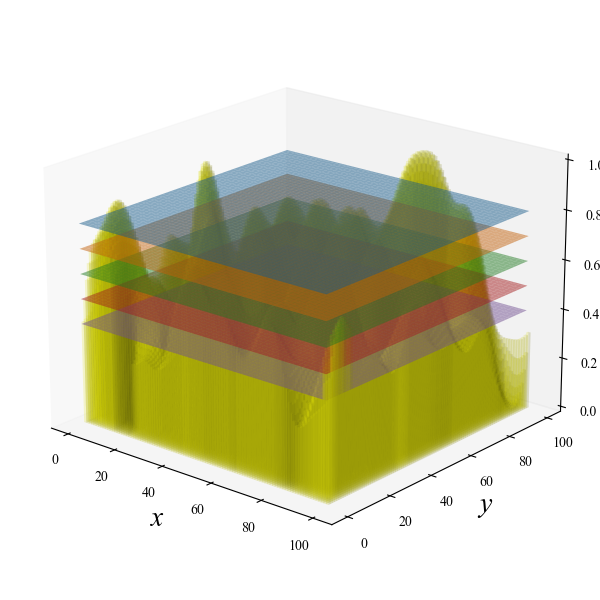}}
  \subfigure[]{\includegraphics[width=0.15\textwidth]{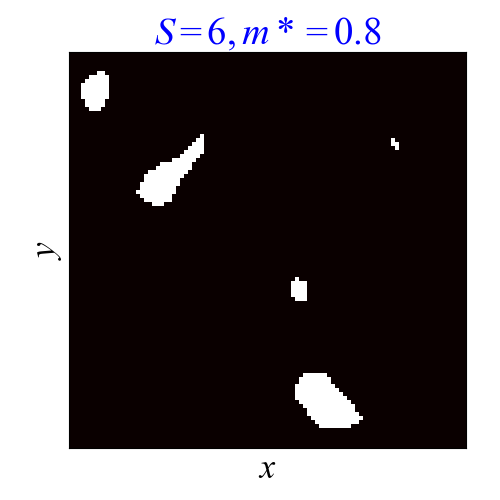}}
  \subfigure[]{\includegraphics[width=0.15\textwidth]{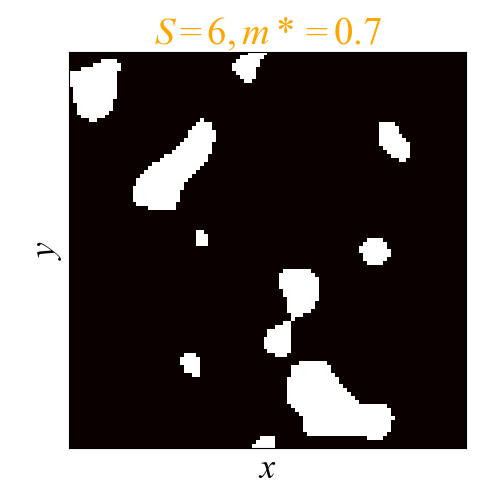}}
  \subfigure[]{\includegraphics[width=0.15\textwidth]{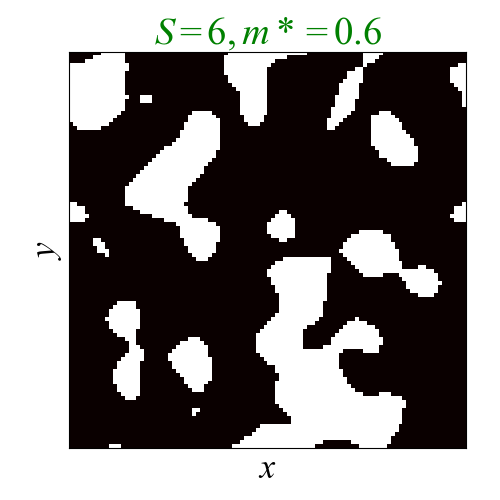}}
  \subfigure[]{\includegraphics[width=0.15\textwidth]{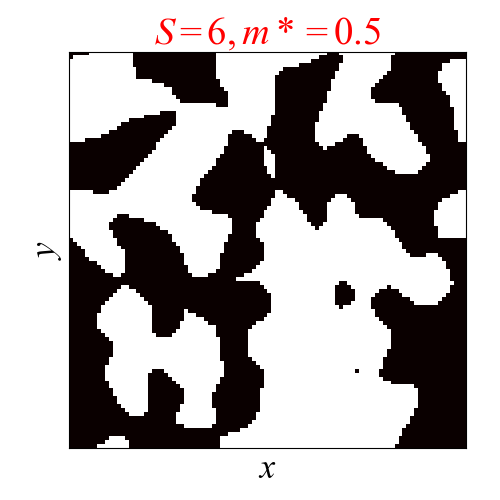}}
  \subfigure[]{\includegraphics[width=0.15\textwidth]{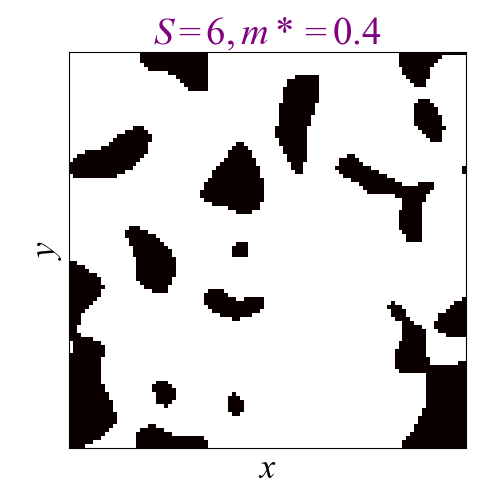}}
  \caption{(a) 3D landscape with smoothness  $S=6$, and $L_x=L_y=100$. (b)--(f) The disordered texture defined by the intersection of the landscape and the ``sea levels'' (b) $m^*=0.8$, (c) $m^*=0.7$, (d) $m^*=0.6$, (e) $m^*=0.5$, and (f) $m^*=0.4$. }
  \label{smfig:S=6}
\end{figure}

\begin{figure}[!htp]
  \centering
  \subfigure[]{\includegraphics[width=0.21\textwidth]{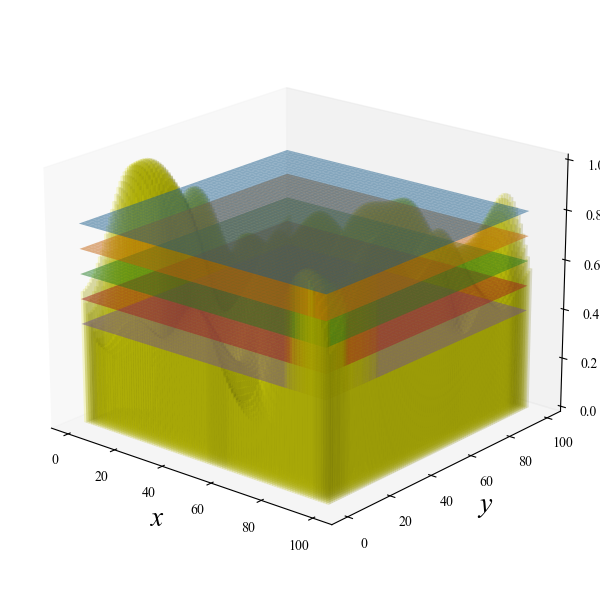}}
  \subfigure[]{\includegraphics[width=0.15\textwidth]{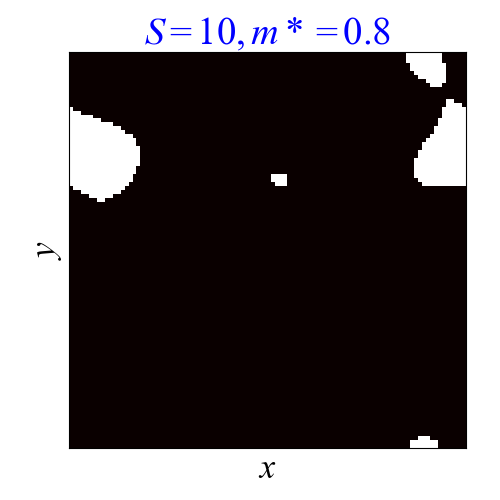}}
  \subfigure[]{\includegraphics[width=0.15\textwidth]{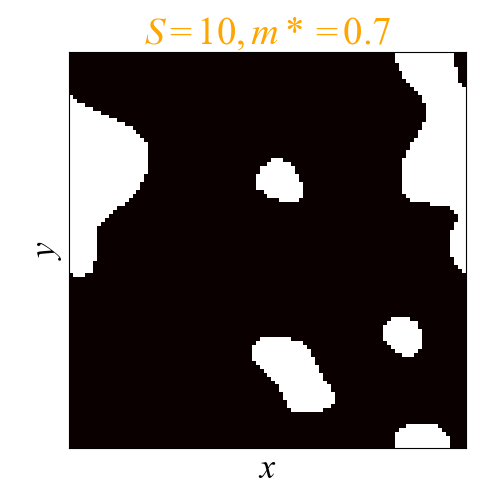}}
  \subfigure[]{\includegraphics[width=0.15\textwidth]{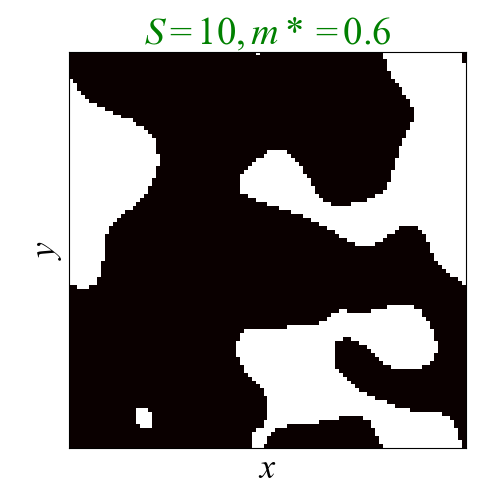}}
  \subfigure[]{\includegraphics[width=0.15\textwidth]{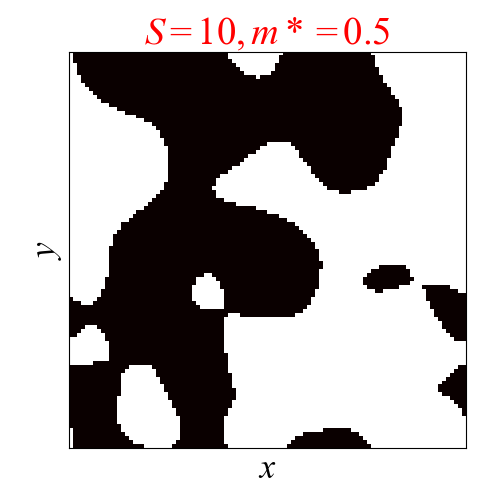}}
  \subfigure[]{\includegraphics[width=0.15\textwidth]{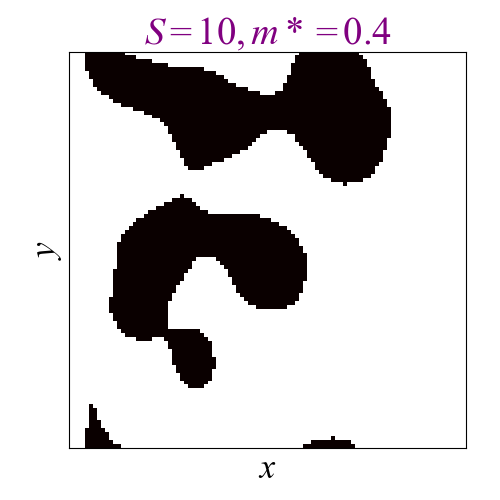}}
  \caption{(a) 3D landscape with smoothness  $S=10$, and $L_x=L_y=100$. (b)--(f) The disordered texture defined by the intersection of the landscape and the ``sea levels'' (b) $m^*=0.8$, (c) $m^*=0.7$, (d) $m^*=0.6$, (e) $m^*=0.5$, and (f) $m^*=0.4$. }
  \label{smfig:S=10}
\end{figure}

For any generic scalar function $f(\bold r)$, one can generate~\cite{Ong2016} a real spatially-correlated random texture $m(\bold r)$ by
multiplying the Fourier transform of $f(\bold r)$ with a random phase, and then taking the real part of the inverse Fourier transform of that:
\begin{equation}
m(\bold r)=\sqrt{2}\,\text{Re}\int d\bold k \int d\bold r'  e^{i\bold k\cdot (\bold r-\bold r') }e^{i\theta(\bold k)}f(\bold r'). \label{smeq:hx}
\end{equation}
Here $\theta(\bold k)$ is drawn from a uniform distribution and is uncorrelated for different $\bold k$. Different choices of $f(\bold r)$ would lead to textures $m(\bold r)$ with different spatial correlation properties. In this work, we specialize $f(\bold r)$ to a Gaussian function in the $x$-$y$ plane: 
\begin{equation}
  f(\bold r) = \frac{1}{\pi S^2}e^{-(x^2+y^2)/S^2}.
\end{equation}
From Eq.~\ref{smeq:hx}, one can show that the spatial correlation in $m(\bold r)$ obeys 
\begin{equation}
  \langle m(\bold r_1) m(\bold r_2) \rangle=\langle m^2\rangle e^{-|\bold r_1-\bold r_2|^2/(2S^2)}.
\end{equation}
$S$, being the correlation length, effectively controls the boundary smoothness of the islands, as shown in subfigures (a) in Figs.~\ref{smfig:S=1}--\ref{smfig:S=10} for $S=1, 2, 3, 6, 10$. For simplicity, $m(\bold r)$ has been linearly shifted to be in the range of $[0,1]$. It is evident that larger values of $S$ result in larger and smoother islands. Based on this random landscape, the disordered topological lattice (with both layers having identical island configurations) is generated with a ``sea level'' $m^*$, such that the topological islands are precisely the regions $\bold r$ where $m(\bold r)\geq m^*$. In analogy to the sea level for an actual geographical landscape, when the sea level is very high, only the peaks above the mountains are visible, and they would form small islands. But when the sea level is lowered, the lower regions emerge from the sea and the islands grow and coalesce into larger islands, eventually forming one large island. 

In Figs.~\ref{smfig:S=1}--\ref{smfig:S=10}(a), the landscapes $m(\bold r)$ are shown inundated by the sea levels $m^*$ (colored in blue, orange, green, red and purple) of different values, corresponding to the island configurations shown in Figs.~\ref{smfig:S=1}--\ref{smfig:S=10}(b)--(f). White regions represent topologically nontrivial islands to be implemented by $\hat H_\text{two-layer}$ with mass $m=1$ (such that the Chern number is $+1$~\cite{kawabata2018anomalous}), while the black areas represent topologically trivial regions which are implemented as empty regions with all bonds removed. To reduce the influence of the edge effects of the system and focus solely on observing the effect caused by the topological islands, periodic boundary conditions are imposed in both the $x$ and $y$ directions. This ensures that the islands smoothly connect across the boundaries.


\subsection{Maximal island width, percolation threshold and critical sea level}

\begin{figure}[!htp]
  \centering
  \subfigure[]{\includegraphics[width=0.38\textwidth]{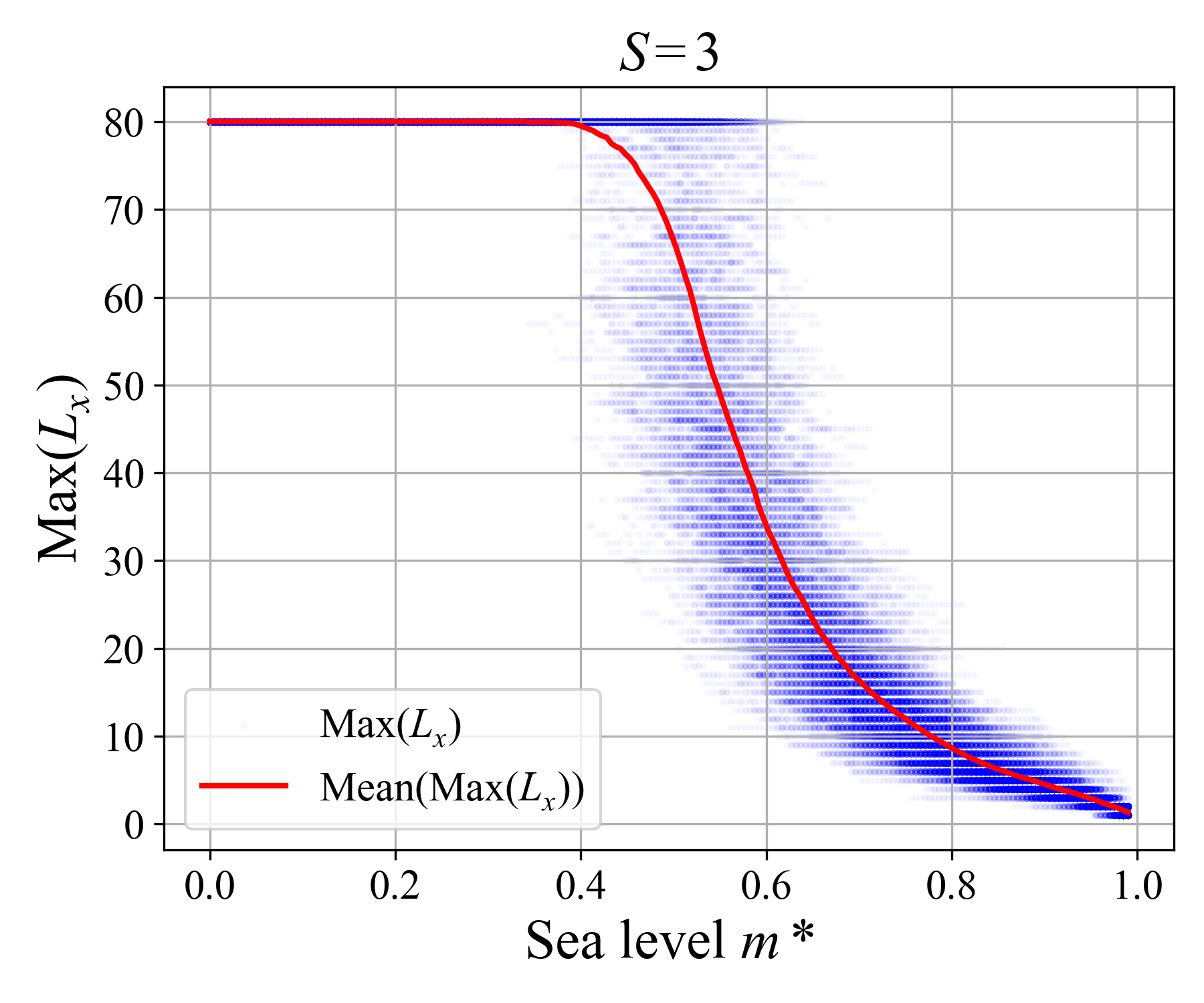}}
  \subfigure[]{\includegraphics[width=0.38\textwidth]{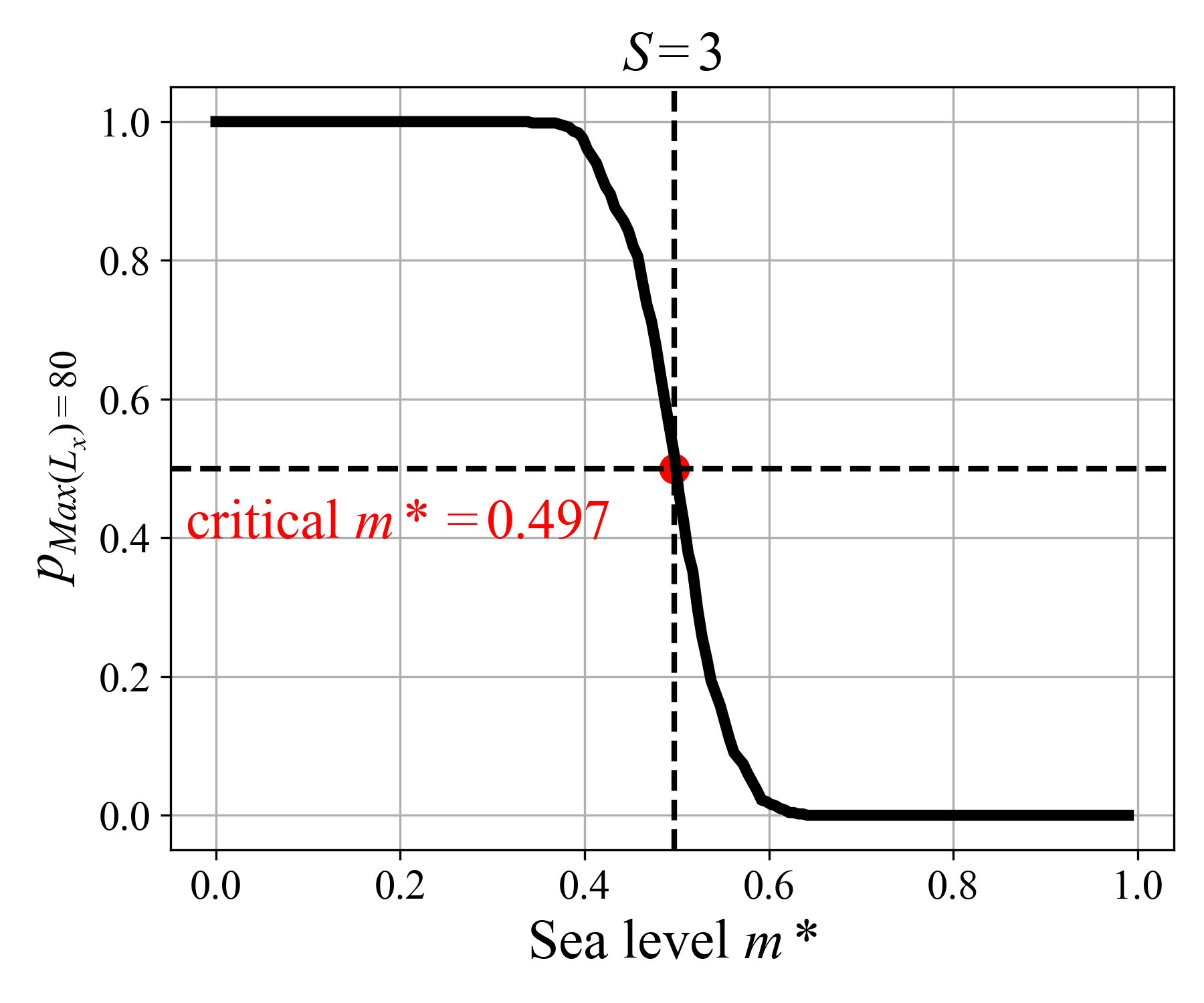}}
  \caption{(a) The maximum island width Max($L_x$) against sea level $m^*\in(0,1)$  (blue dots) for systems with $80\times 30$ unit cells. Darker colors represent a higher probability. The mean value of $L_x$ is shown as the red line. (b) The crossing probability $p_{\text{Max}(L_x)=80}$ across the whole system, against sea level $m^*$. The critical $m^*$ is taken to be at the middle $(P_{\text{Max}(L_x)=80}=0.5)$ of the crossing probability curve, which is $0.497$ for this $S=3$ system. The number of samples for each $m^*$ is 500.
  }
  \label{smfig:percolation_threshold}
\end{figure}

We generate a number of random lattices of size $L_x=80$ and $L_y=30$ with edge smoothness  $S=3$ using the method described above in the Supplemental Material. We plot the possible values of Max($L_x$), which is the maximum horizontal width of the topological island, against the sea level $m^*$ in Fig.~\ref{smfig:percolation_threshold}(a). The red line shows the mean value of Max($L_x$) for each $m^*$. We see that while $m^*\approx 1$ gives extremely tiny islands, Max($L_x$) rapidly increases as $m^*$ decreases, and soon saturates at the maximal system width of 80 unit cells as the islands critically coalesce.

We also plot the crossing probability $p_{\text{Max}(L_x)=80}$, which is the probability that the percolating island spans the entire lattice horizontally, against $m^*$ in Fig.~\ref{smfig:percolation_threshold}(b). Note that for topologically guided gain to occur, we just need the islands to coalesce enough such that  $\text{Max}(L_x)>x^*$ i.e. percolate across a width of $x^*$, and not necessarily the whole system. We define the critical $m^*$ (also not to be confused with the critical width $\text{Max}(L_x)=x^*$ for an island to start supporting topologically guided gain) as the value where the crossing probability is $0.5$, which is marked by the red circle. For $S=3$, we find that the critical $m^*$ is $0.497$, which is remarkably close to that of the square lattice percolation threshold, despite our islands being irregularly shaped.

\section{Dynamics in our disorder bilayer non-Hermitian Chern islands}

\begin{figure}[!htp]
  \centering
 \includegraphics[width=0.7\textwidth]{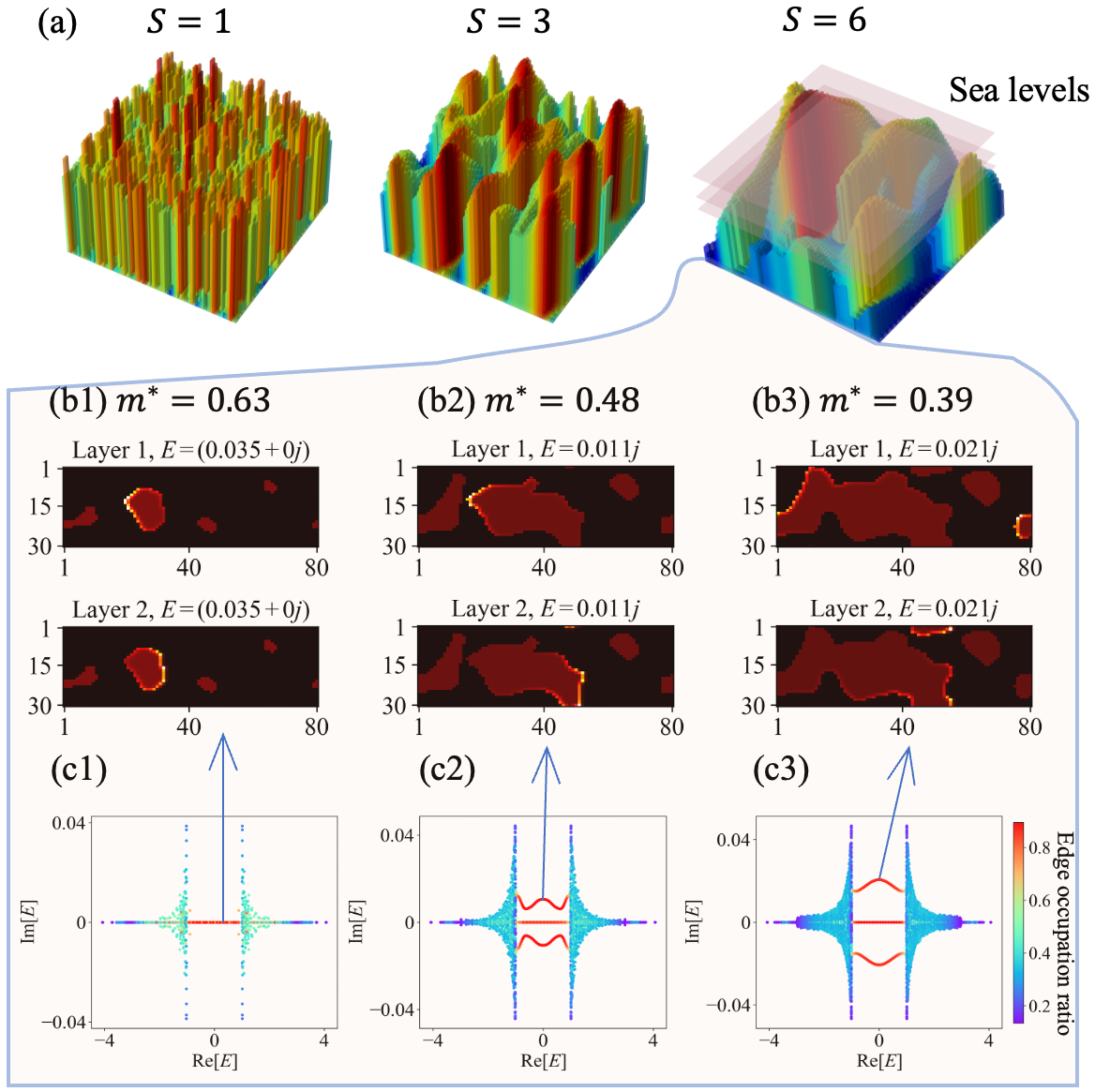}
  \caption{(a) $m(\bold r)$ landscapes which become smoother as the edge smoothness  $S$ increases. (b1)--(b3) Illustrative resultant island configurations with midgap edge states highlighted by spatial intensity. Other dark red islands that do not have highlighted edges do not contain edge states. (c1)--(c3) Energy spectra colored by edge occupation ratio according to Eq.~(4) from the main text, for the landscapes in (b1)--(b3) respectively. Here, we have set $\mu_0=0.001$ and $\gamma=\pm0.05$. 
  }
  \label{fig:land_generation}
\end{figure}

Fig.~\ref{fig:land_generation}(a) provides illustrative examples of the landscapes and islands generated by our approach above, with larger $S$ clearly leading to smoother landscapes. Figs.~\ref{fig:land_generation}(b) depict three distinct landscapes corresponding to sea levels of $m^*=0.63, 0.48$, and $0.39$. The spatial profiles of their midgap ($\text{Re}[E]=0$) eigenstates are shown in Fig.~\ref{fig:land_generation}(b), where we clearly see that they are confined along the islands' boundaries. Spectra colored by edge occupation ratio are shown in Fig.~\ref{fig:land_generation}(c). The edge occupation ratio is defined in the main text Eq.~(4). We observe PT symmetry breaking, characterized by the emergence of complex energies for topological edge states, in Figs.~\ref{fig:land_generation}(c2) and (c3), where the islands have grown large enough such that their maximum width Max($L_x$) is sufficiently large.

\subsection{Effect of the nonsmoothness of island boundaries on topological bulk leakage}

In the main text, we have mentioned that the edge dynamics involves a competition between topologically guided propagation/gain and bulk leakage, which can lead to unlimited growth whose path is not topologically controlled. Therefore, the island boundaries cannot be excessively sharp (small $S$), which can overshadow the edge PT transition with bulk leakage. 
Here, we further show that the smoothness of the island boundaries is a key factor in determining the topological bulk leakage.

\begin{figure}[!htp]
  \centering
  \subfigure[]{\includegraphics[width=0.26\textwidth]{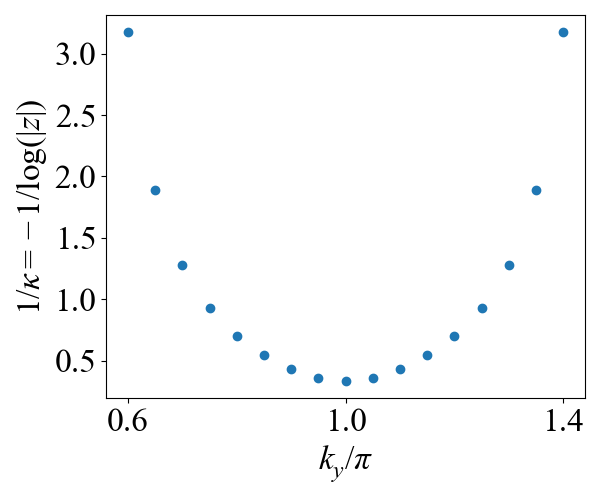}}
  \subfigure[]{\includegraphics[width=0.26\textwidth]{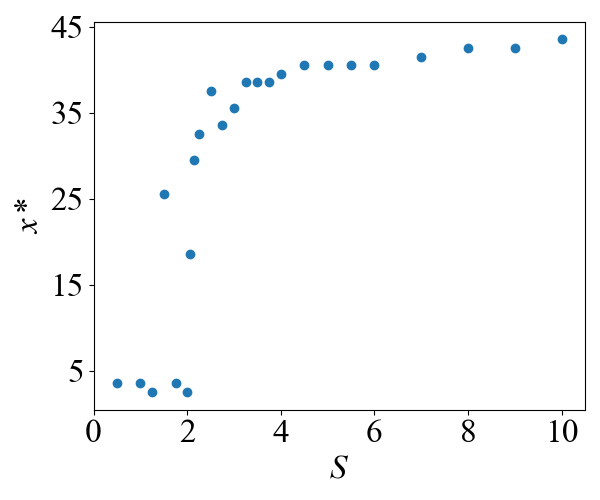}}
  \subfigure[]{\includegraphics[width=0.26\textwidth]{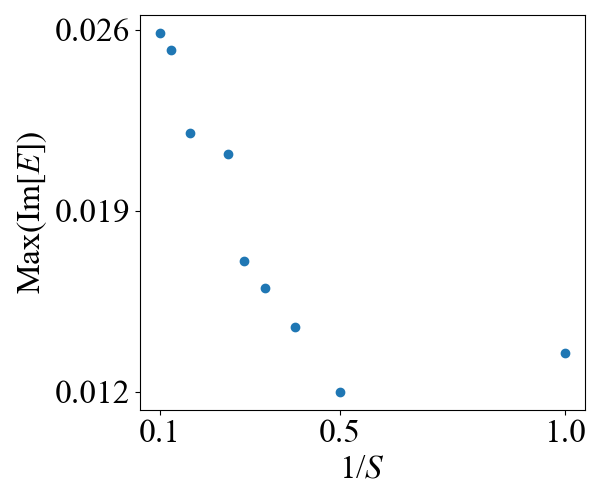}}
  \subfigure[]{\includegraphics[width=0.27\textwidth]{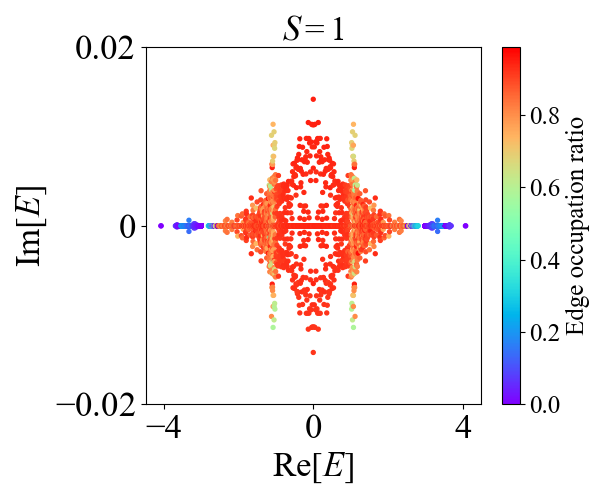}}
  \subfigure[]{\includegraphics[width=0.27\textwidth]{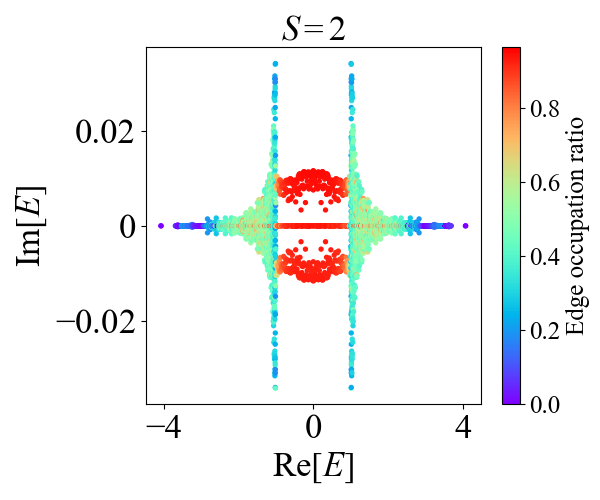}}
  \subfigure[]{\includegraphics[width=0.27\textwidth]{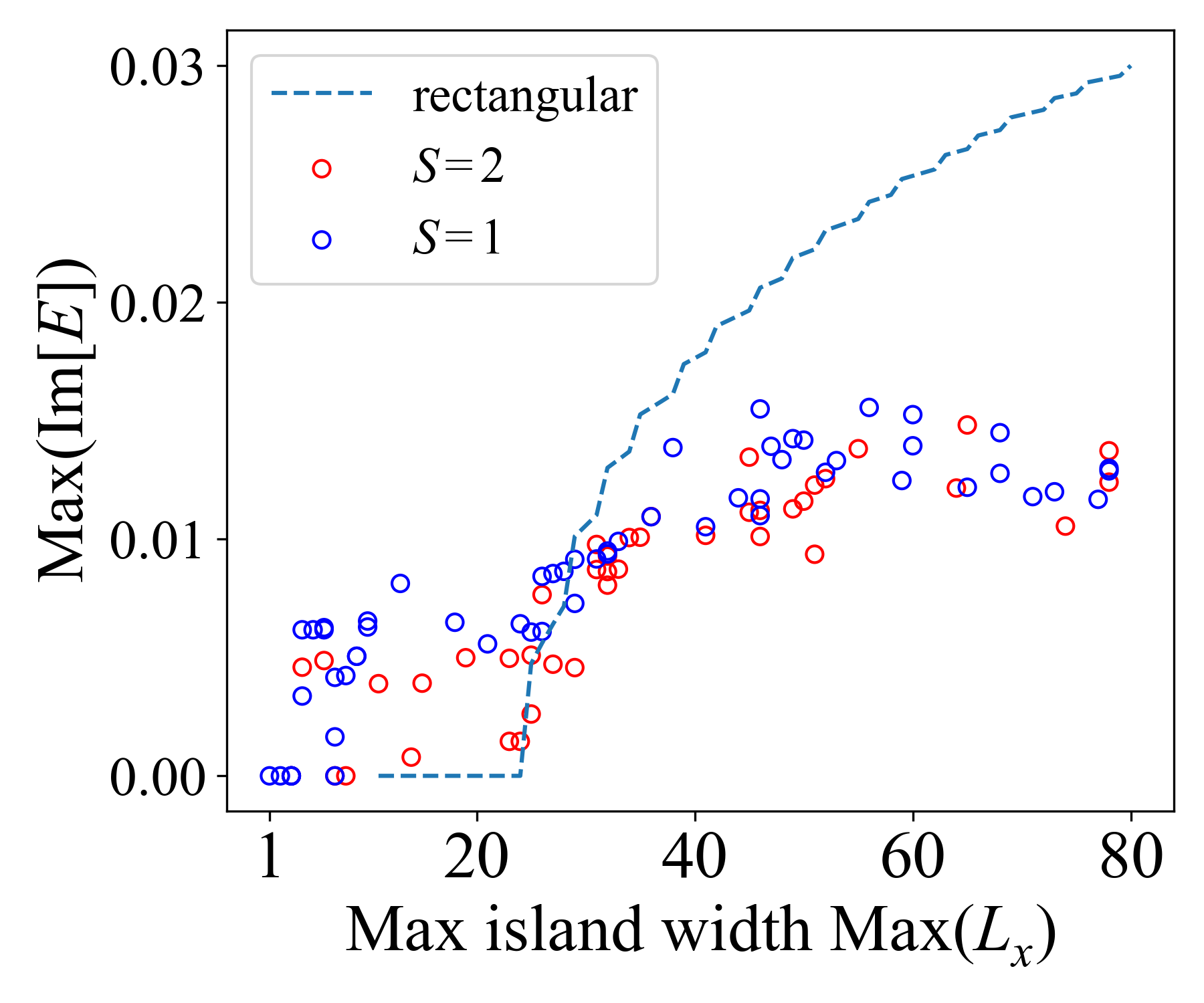}}
  \caption{(a) Spatial state decay length $1/\kappa$ against $k_y$ for our model $\hat H_{two-layer}$, which diverges as $k_y$ enters the bulk. (b) Critical width $x^*$ for growth as a function of island boundary smoothness $S$, which saturates at high smoothness $S$. (c) Max(Im[$E$]) increases with $S$ for topological edge states. (d) and (e) Energy spectra for $S=1$ and $2$ respectively, for a $50\times 30$ unit cell system. (f) For $S=1$ and $2$, the absence of a PT transition in Max(Im[$E$]) for midgap topological edge states as the maximum island width Max($L_x$) is varied, while we keep $L_y=30$ fixed.	All plots are generated with parameters $\mu_0=0.005, \gamma=0.05$, and (a)--(c) are generated on a $80\times 30$ lattice.
	} 
  \label{smfig:bulkleak}
\end{figure}

In Fig.~\ref{smfig:bulkleak}(a), we show that the spatial decay length $1/\kappa$ of the in-gap topological modes of our model $\hat H_\text{two-layer}$ diverge as we enter the bulk. For any 
eigenstate with energy $E$, we can calculate its decay length $1/\kappa$ by solving Det$[H(z, k_y)-E \mathbb{I}]=0$ \cite{lee2019anatomy,yao2018edge,lee2020unraveling}, where $z=e^{ik_x-\kappa}$. As shown in Fig.~\ref{smfig:bulkleak}, at the center of the energy gap, i.e., $k_y=\pi$, $1/\kappa$ is small, but as one goes closer to the bulk states near $k_y=0.6\pi$ and $1.4\pi$, the decay depth gets longer as the eigenstate merges with the bulk. Hence as a system is dynamically driven, states with larger and larger decay lengths would be excited, and eventually, we excite those which are comparable to the feature size of the islands. Henceforth, we start to see significant bulk leakage.


In Fig.~\ref{smfig:bulkleak}(b), we see that the critical island width $x^*$ for gainy behavior is very small for small $S$ (sharp island features), but rapidly increases with $S$ till it saturates at large $S$ (smooth island boundaries). This is because with very small islands features characterized by small radii of curvature and inter-island distance, bulk leakage is almost imminent, and bulk eigenstates have broken PT symmetry. As island features become larger with increasing $S$, bulk leakage becomes drastically reduced with the island features having a characteristic length scale exceeding that of $1/\kappa$ of the in-gap topological modes (Fig.~\ref{smfig:bulkleak}(a)). As bulk leakage is minimized, the remaining growth mechanism is topologically guided gain, and that fixes $x^*$ at the constant value derived in the main text. However, even though $x^*$ effectively saturates at large $S$, the actual growth rate max(Im[$E$]) from the edge states continues to increase as $S$ increases (i.e. as $1/S$ decreases), as shown in Fig.~\ref{smfig:bulkleak}(c). This is because smoother island boundaries allow for more undisturbed topological propagation without destructive interference, and therefore topologically guided gain. 

In Figs.~\ref{smfig:bulkleak}(d) and (e), we show the full spectra for the $S=1$ and $S=2$ cases, which are the cases dominated by bulk leakage. For $S=1$, the island features are so tiny that there is no proper distinction between the bulk and the boundaries (with most states all red). For $S=2$, the bulk states (greenish blue) are clearly distinguishable from the boundary states (red), but the boundary states do not form well-defined chiral bands and the bulk states have far larger gain (Im[$E$]). Hence the inevitably significant bulk leakage for small $S=1$ or $2$.


To more concretely show how that bulk leakage is related to island width, we plot the maximum Im[$E$] values of midgap topological edge states against the maximum island width Max($L_x$) in Fig.~\ref{smfig:bulkleak}(f). The maximum Im[$E$] values of the topological edge states are indicated by empty circles, corresponding to boundary smoothness $S= 1$ or $2$. One does not observe a similar PT transition as those in Fig.~3(a) in the main text. The reason is that the edges are too sharp to allow for well-defined chiral Chern dynamics, such that bulk leakage dominates.

As an additional comment, bulk leakage can become much more significant if the system is too narrow, such that the islands are also inevitably narrow and interference can occur between two opposite boundaries of the island.

\color{black}
\section{The relation between the percolation transition and the topologically guided gain}

When it comes to PT symmetry breaking, the EP scaling is often discussed. Here, we thus provide more details. In the following, we will firstly explain that the EP scaling is not directly related to our tologically guided gain model, and then we will discuss the relation between the percolation transition and the topologically guided gain. 

Concerning scaling relations near the EP, the conventional critical percolation networks typically analyze the power-law decay of 2-point functions, focusing on the ``bulk'' sites within the islands. However, our discussion here primarily revolves around the topological edge dynamics occurring at the boundaries of these islands. 

Although the critical scaling behaviors are not directly linked, what remains pertinent is that the topologically guided gain ultimately depends on the width of the islands. This dependency arises because non-Hermitian gain is highly sensitive to variations in island widths, except when influenced by scattering from disorder and sharp corners (where we have included the content about the sharp corners in the Supplementary). Consequently, at the percolation transition, there is a notable shift in the crossing probability, leading to a qualitative transformation in topologically guided gain (manifested as the EP or real-complex $E$ transition) when the widest island surpasses a certain threshold in growth. Specifically, the functional dependence between the system parameter and the largest island size encodes information about the imaginary part of $E$ for the topologically guided gain model.

\section{Possible experiments proposal of our percolation model in circuits implementation}

In our main text, we have theoretically demonstrated the percolation-induced PT transition in a bilayer non-Hermitian Chern insulator. Our model and topological guided gain mechanism does not require many-body interactions, and can thus be realized in both classical and quantum platforms. Appropriately designed metamaterial platforms and quantum simulators are suitable candidates in principle, as long as they can provide the three requisite physical ingredients: (i) non-Hermitian pumping; (ii) bilayer structure with interlayer tunneling and oppositely directed pumping; and (iii) chiral topological pumping. (i) has been realized in electrical circuits~\cite{helbig2020generalized,liu2021non,zou2021observation}, photonic quantum walks~\cite{weidemann2020topological,xiao2020non,wang2021generating}, mechanical metamaterials~\cite{ghatak2020observation}, ultracold atoms~\cite{li2020topological,liang2022dynamic} and quantum circuits~\cite{smith2019simulating,gou2020tunable,koh2022simulation,kirmani2022probing,frey2022realization,chertkov2023characterizing,chen2023robust,yang2023simulating,iqbal2023creation,shen2023observation,koh2023observation}. (ii) can typically be realized as long as the platform can be made to support coupled bilayers, which is definitely feasible in electrical circuits~\cite{hofmann2019chiral,ezawa2019electric,helbig2020generalized,hofmann2020reciprocal,liu2020gain,liu2021non,stegmaier2021topological,zhang2020non,zhang2022observation,shang2022experimental,yuan2023non,zou2023experimental,zhu2023higher,zhang2023electrical}, suitable photonic crystals~\cite{pan2018photonic,xiao2020non,zhu2020photonic,song2020two,ao2020topological} and quantum circuits~\cite{smith2019simulating,gou2020tunable,koh2022simulation,kirmani2022probing,frey2022realization,chertkov2023characterizing,chen2023robust,yang2023simulating,iqbal2023creation,shen2023observation,koh2023observation}. (iii) has also been realized in various photonic~\cite{weidemann2020topological}, mechanical~\cite{ghatak2020observation} and acoustic systems~\cite{zhang2021acoustic}, as well as in cold atoms~\cite{liang2022dynamic} and quantum circuits~\cite{roushan2014observation}. As such, suitably designed photonic systems, electrical circuits and quantum circuits on digital quantum processors can realize our model and percolation-induced PT breaking just with existing experimental technology.

While appropriately designed metamaterial platforms are in-principle all feasible, we recommend using electrical circuits due to their versatility~\cite{lee2018topolectrical,hofmann2019chiral}. In general, it is feasible to construct a tailored topolectrical circuit to simulate a non-Hermitian Chern insulator for each layer, and subsequently couple them through mutual inductances. By doing so, we can detect the percolation-induced PT transition by analyzing the impedance across the entire circuit. Below we provide a brief proposal of how our model can be realized and observed in such a system.

Detailed in Hofmann and Helbig et al.'s work~\cite{hofmann2019chiral} are experimentally accessible circuits for Chern lattices. They propose achieving an effective Chern lattice by breaking reciprocity and time-reversal symmetry using negative impedance converters with current inversion (INICs), which are based on operational amplifiers (OpAmps). In Ref.~\cite{hofmann2019chiral}, the INICs are perfectly matched so that the effective model Laplacian is Hermitian. However, we can simply replace the effective resistances in their setup with dissimilar resistances $R_{\text{a}}\neq R_{\text{b}}$ to introduce non-Hermiticity. In the following, we will introduce: ``INIC circuit with unequal internal resistances $\nu=R_{\text{a}}/ R_{\text{b}}\neq 1$'' and how to construct a Chern model with non-Hermiticity by choosing two different $\nu_A$ and $\nu_B$ for two oppositely circular sets of INICs.


\subsection{Description of INICs -- the building blocks of a Chern circuit}

The topological nature of the model is rooted in the INIC next-nearest neighbor $A$-$A$ and $B$-$B$ coupling elements. These elements break both time-reversal symmetry and circuit reciprocity by effectively implementing negative and positive resistance in the forward and reverse directions of the element, respectively. We adjust the resistance ratios, denoted by $\nu_A$ and $\nu_B$, of the INICs in the $A$-$A$ (highlighted in red in Fig.~\ref{rfig:topoelec_chernb}(b)) and $B$-$B$ (highlighted in blue in Fig.~\ref{rfig:topoelec_chernb}(b)) coupling elements.

\begin{figure}[!htp]
    \centering
    \includegraphics[width=0.7\textwidth]{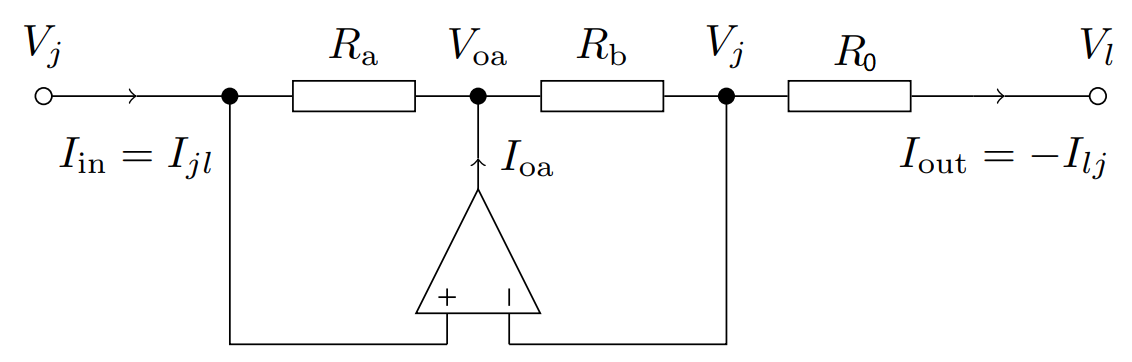}
    \caption{\textcolor{black}{A sketch of an INIC circuit, adapted from Supplemental Material of Ref.~\cite{hofmann2019chiral}.}}
    \label{rfig:INICb}
\end{figure}

\begin{figure}
    \centering
    \includegraphics[width=0.96\textwidth]{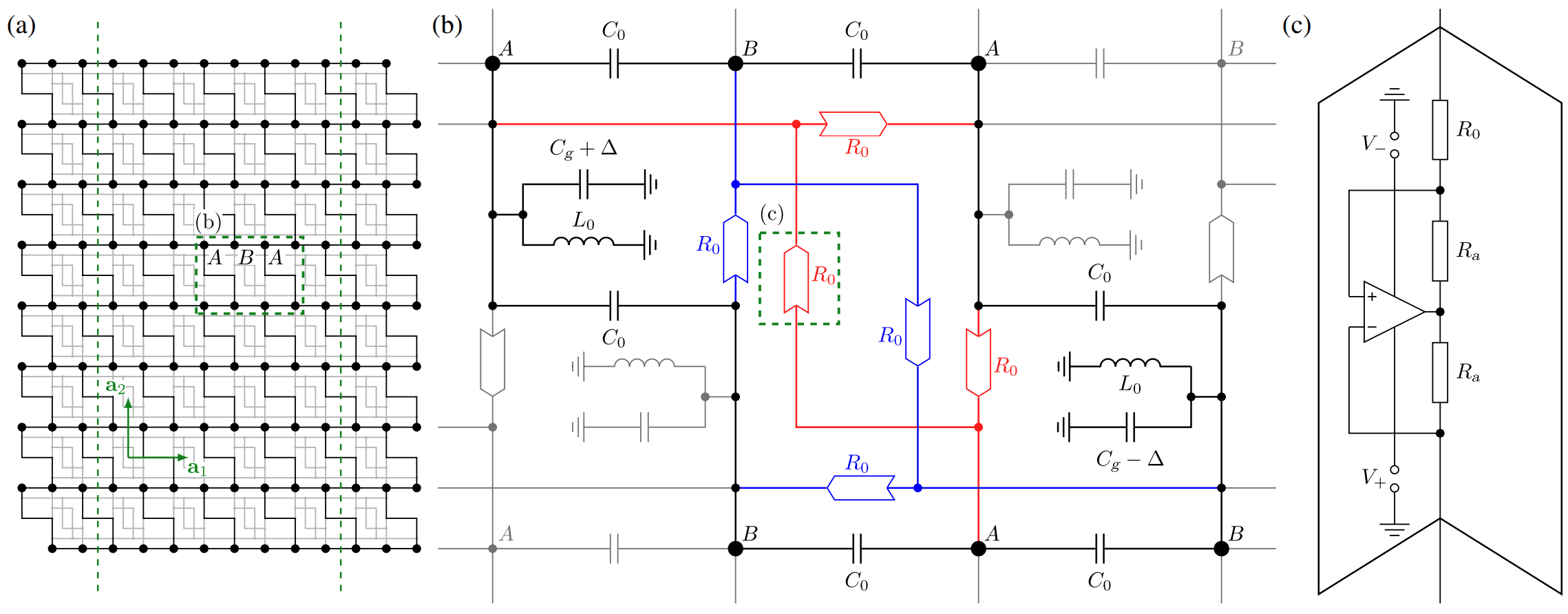}
    \caption{\textcolor{black}{Topoelectrical Chern circuits, adapted from Ref.~\cite{hofmann2019chiral}. (a) A sketch of the topoelectrical Chern circuit. (b) The circuit unit cell consists of two nodes, each of which is connected to three adjacent nodes through a capacitor $C_0$ and to six next-nearest neighbors through INICs. (c) INICs.}}
    \label{rfig:topoelec_chernb}
\end{figure}

As shown in Fig.~\ref{rfig:INICb}, the current entering the INIC from the left, denoted as $I_{\text{in}}$, is determined by $I_{\text{in}}=(V_j-V_{\text{oa}})/R_{\text{a}}$, while the outgoing current on the right, denoted as $I_{\text{out}}$, is given by $I_{\text{out}}=(V_j-V_l)/R_0$ when the OpAmp operates in a negative feedback configuration. Assuming infinite impedance at the OpAmp inputs which prevents any current flowing into the OpAmp, the output current is simply $I_{\text{out}}=(V_{\text{oa}}-V_j)/R_{\text{b}}$. Solving these equations for $I_{\text{in}}$ results in

\begin{equation}
\begin{aligned}
& I_{\text{in}}=-\frac{R_{\text{b}}}{R_{\text{a}} \cdot R_0}(V_j-V_l), \\
& I_{\text{out}}=\frac{1}{R_0}(V_j-V_l).
\end{aligned}
\end{equation}
Expressing these outcomes in Laplacian form yields the node voltage equation
\begin{equation}
\begin{pmatrix}
I_{j l} \\
I_{l j}
\end{pmatrix}=\frac{1}{R_0}\begin{pmatrix}
-\nu & \nu \\
-1 & 1
\end{pmatrix}\begin{pmatrix}
V_j \\
V_l
\end{pmatrix},
\end{equation}
where $\nu=R_{\text{b}}/R_{\text{a}}$. Notably, the matrix is asymmetric and non-Hermitian, indicating the breaking of circuit reciprocity.


\subsection{Constructing a Chern model with non-Hermiticity by choosing two different $\nu_A$ and $\nu_B$ for two oppositely circular sets of INICs}

\color{black}
Below, we provide a brief proposal of how our model can be realized and observed in such a system. As shown in Ref.~\cite{hofmann2019chiral}, the topoelectrical chern circuit is formed by a periodic circuit structure sketched in Fig.~\ref{rfig:topoelec_chernb}(a). It resembles to the Haldane model, inducing an additional effective magnetic flux, thus giving the nontrivial Chern number.

Let us first clarify the connection between the circuit Laplacian and conventional Hamiltonian in the following. It is worth noting that the performance of an RLC circuit is dictated by its circuit Laplacian, akin to how the Hamiltonian describes the dynamics of a physical system; however, it is a convention to use the Laplacian in circuit theory. 

We define voltage and current vectors by denoting the voltages measured at the nodes of a circuit board against ground as well as the input currents at the nodes. These are represented by $N$-component vectors $\bold V$ and $\bold I$, respectively. The motion of the electrical circuit is 
\begin{equation}
\frac{d}{d t} \mathbf{I}(t)=C \frac{d^2}{d t^2} \mathbf{V}(t)+\Sigma \frac{d}{d t} \mathbf{V}(t)+L \mathbf{V}(t),
\end{equation}
where capacitance $C$, conductance $\Sigma$, and inductance $L$ are the real-valued $(N \times N)$-matrices forming the grounded circuit Laplacian by
\begin{equation}
J(\omega)=i \omega C+\Sigma+\frac{1}{i \omega} L .
\end{equation}

The homogeneous equations of motion $(\mathbf{I}=0$, where the circuit's time evolution is solely determined by its eigenfrequencies) can be re-written as $2 N$ differential equations of first order
\begin{equation}
i \frac{d}{d t} \psi(t)=H \psi(t),
\end{equation}
where $\psi(t):=(\dot{\mathbf{V}}(t), \mathbf{V}(t))^{\top}$. The time evolution of a given eigenstate $\psi_\alpha(t)$ yields $\psi_\alpha(t)=\psi_\alpha e^{-i \omega_\alpha t}$ where the eigenvalues $\omega_\alpha, \alpha \in$ $\{1, \ldots, 2 N\}$ are the resonance frequencies of the system shown in Fig.~\ref{rfig:topoelec_chernb}, defined as the roots of the admittance eigenvalues $j\left(\omega_\alpha\right)=0$. It means that for the circuits, we can simply obtain the Laplacian $J$ by measuring the impedance of the circuit, and the eigenspectrum can be obtained by exact diagonalization of the Laplacian. 

The grounded circuit Laplacian $J$ is defined as the matrix relating the vector of voltages $\bold V$ measured with respect to ground to the vector of input currents $\bold I$ at the $N$ circuit nodes by $\bold I=J\bold V$. For an ac frequency $\omega=2\pi f$ and two-dimensional reciprocal space implied by the brick wall gauge, the topological Chern circuits (TCC) Laplacian $J_{\mathrm{TCC}}(\mathbf{k}; \omega)$ is given by
\begin{equation}
    J_{\mathrm{TCC}}(\mathbf{k}; \omega) = i\omega \left(J_0\mathbf{1}+J_x\sigma_x+J_y\sigma_y+J_z\sigma_z\right). 
\end{equation}
where the parameters $J_0, J_x, J_y, J_z$ are given by the following equations. $C_0, C_g, L_0, R_0, \Delta$ are indicated in the Fig.~\ref{rfig:topoelec_chernb}. The corresponding components are 
\begin{equation}
    J_0=3 C_0+C_g-\frac{1}{\omega^2 L_0}-i\frac{1}{\omega R_0}(1-\frac{\nu_A+\nu_B}{2})\left[3-\cos \left(k_x\right)-\cos \left(k_y\right) - \cos \left(k_x-k_y\right)\right].
\end{equation}

\begin{equation}
    J_x=-C_0\left[1+\cos \left(k_x\right)+\cos \left(k_y\right)\right].
\end{equation}

\begin{equation}
    J_y=-C_0\left[\sin \left(k_x\right)+\sin \left(k_y\right)\right].
\end{equation}

\begin{equation}
    \begin{aligned}
    J_z= \Delta&+\frac{1}{\omega R_0}\left(1+\frac{\nu_A+\nu_B}{2}\right)\left[\sin \left(k_x\right)-\sin \left(k_y\right)-\sin \left(k_x-k_y\right)\right]\\
    &+i\frac{1}{\omega R_0}\frac{\nu_A-\nu_B}{2}\left[3-\cos \left(k_x\right)-\cos \left(k_y\right) - \cos \left(k_x-k_y\right)\right].
    \end{aligned}\label{eq:Jzb}
\end{equation}

We notice that, for the perfectly matched INICs, i.e., $\nu_A=\nu_B=1$, the effective Laplacian is Hermitian, and is reduced to the Eq. (1) in Ref.~\cite{hofmann2019chiral}. However, by introducing a slight difference between $\nu_A\neq1$ and $\nu_B\neq1$ ($\nu_A\neq\nu_B$), the effective Laplacian becomes non-Hermitian, as indicated in the last term in Eq.~\ref{eq:Jzb}, leading to the breaking of circuit reciprocity.

\color{black}

\textcolor{black}{Now that we have two non-Hermitian layers with nontrivial Chern number, we can couple two layers.} Weak coupling between the two layers can be effective through the mutual inductance between the two layers. By removing some of the nodes in the circuit (at the same spot in each layer), we can simulate the percolation islands.

\color{black}
Since the wavefunction is physically represented as the electrical potential, it can be dynamically mapped out in real-time by continuously measuring the voltage at each desired node~\cite{hofmann2019chiral,helbig2020generalized,helbig2019band}. The PT transition, which is marked by the departure of the circuit Laplacian edge eigenvalues away from the real line (i.e. from Im$(E)=0$ to Im$(E)\neq 0$ at Re$(E)=0$), can also be detected by measuring the topolectrical impedance~\cite{liu2020octupole,song2020realization,liu2021non} between any two nodes, which exhibits a resonance peak only when a particular chosen eigenvalue is present.


\textcolor{black}{It is crucial to underscore that our decision to present an experimental proposal within the context of a topoelectrical circuit stems from the inherent versatility of these circuits. Their adaptability allows for the simulation of various models~\cite{lee2018topolectrical,imhof2018topolectrical,hofmann2019chiral}. In the discussion above, we employed a Chern topoelectrical circuit as a model for our topologically guided gain. While we have not provided a circuit model identical to our topologically guided gain model, it is important to note that the concept of topologically guided gain necessitates only a nontrivial Chern number. This highlights the flexibility and broad applicability of topoelectrical circuits in this domain. }


\subsection{The physical units of the circuit parameters}

Since we have discussed a feasible experimental proposal, let us use topoelectrical circuits as an example. We have previously derived the Laplacian matrix of the topoelectrical circuits, given by $J_{\mathrm{TCC}}(\mathbf{k}; \omega) = i\omega \left(J_0\mathbf{1}+J_x\sigma_x+J_y\sigma_y+J_z\sigma_z\right)$, with each component detailed in the response to the previous question. Now, in the main text, we utilize a non-Hermitian Chern insulator, represented by $H_{\text{Ch}}(\bold k)=(m+t\cos k_x+t\cos k_y)\sigma_x+(i\gamma+t\sin ky)\sigma_y+t\sin k_x\sigma_z$. In the proposed circuit model, the lossy term resides in the $\sigma_z$ component, while in the Chern model in our main text, the non-Hermiticity is in the $\sigma_y$ component. To gain a rough understanding of the physical units and magnitudes of the parameters, we can compare each component after performing a basis change $H_{\text{Ch}}^\prime(\bold k)=UH_{\text{Ch}}(\bold k)U^\dagger$. Given the basis change defined as
\begin{equation}
U=\frac{1}{\sqrt{2}}\begin{pmatrix}
1 & -i \\
i & -1
\end{pmatrix}, \quad U\sigma_x U^\dagger=-\sigma_x, \quad U\sigma_y U^\dagger=\sigma_z, \quad U\sigma_z U^\dagger=\sigma_y.
\end{equation}
We obtain $H_{\text{Ch}}^\prime(\bold k)=-(m+t\cos k_x+t\cos k_y)\sigma_x+t\sin k_x\sigma_y+(i\gamma+t\sin ky)\sigma_z$. Then, we can roughly compare each component in the following equations
\begin{equation}
\begin{aligned}
J_x&=-C_0\left[1+\cos \left(k_x\right)+\cos \left(k_y\right)\right]\\ &\rightarrow -(m+t\cos k_x+t\cos k_y),\\
J_y&=-C_0\left[\sin \left(k_x\right)+\sin \left(k_y\right)\right]\\&\rightarrow t\sin k_x,
\end{aligned}
\end{equation}
\begin{equation}
\begin{aligned}
J_z&= \Delta+\frac{1}{\omega R_0}\left(1+\frac{\nu_A+\nu_B}{2}\right)\left[\sin \left(k_x\right)-\sin \left(k_y\right)-\sin \left(k_x-k_y\right)\right]\\
&+i\frac{1}{\omega R_0}\frac{\nu_A-\nu_B}{2}\left[3-\cos \left(k_x\right)-\cos \left(k_y\right) - \cos \left(k_x-k_y\right)\right]\\
&\rightarrow i\gamma+t\sin k_y
\end{aligned}
\end{equation}
Though the physical units of the parameters depend on the specific experimental platform, we can gain a sense of the magnitudes of the parameters. It would be helpful if we set 
\begin{equation}
\begin{aligned}
m&=t=C_0,\\
\gamma&=\frac{1}{\omega R_0}\frac{\nu_A-\nu_B}{2}.
\end{aligned}
\end{equation}

In the circuits, the physical units of of $m$ and $t$ is in the unit of capacitance, while the physical unit of $\gamma$ is in the unit of the inverse of $\omega R_0$, which is also in the unit of capacitance. All the parameters are very flexible and can be adjusted in the experiment. For example, here is a possible set of parameters. In the Chern circuits, we can include the operational amplifiers LT1363, and let $m=t=C_0=0.1\mu$F, then $\gamma=0.05*0.1\mu$F$=0.005\mu$F$=\frac{1}{\omega R_0}\frac{\nu_A-\nu_B}{2}$, which can correspond to $\nu_A=1.2,\nu_B=1.1,R_0=100\Omega$, and $\omega=100$kHz. 

\section{The comparison of our PT transition mechanism with the conventional PT transition mechanism}

Our PT symmetry breaking mechanism from topologically guided gain is beyond the usual PT breaking intrinsic to the medium, in the absence of any of the three ingredients such as interlayer coupling.  We shall elaborate on that in two parts: Firstly, we show that if the coupling $\mu_0=0$, the topological spectrum for the bilayer system is fully real. It is the coupling $\mu_0$ that induce such PT transition. Secondly, we compare the mechanism of PT transition in our model with the conventional PT transition mechanism. We show that in our model, the PT transition is not triggered by the gain/loss parameter $\kappa$ exceeding a certain threshold, but by the width of the spatial islands - topologically guided gain.

\color{black}

It is worth highlighting that even with a very faint gain/loss, typically insufficient to trigger a PT transition in other contexts, our topologically guided gain mechanism (assisted by topological pumping) could still facilitate symmetry breaking, provided the islands coalesce and grow sufficiently. This is a novel and surprising aspect of our work.

\setcounter{subsection}{0}
\subsection{Topological spectrum is real in the bilayer system with $\mu_0=0$}

We fully agree with the referee's statement that ``losses and gain do break the time-reversal symmetry, amplification mechanisms that yield a net imaginary part associated to amplification are well-known in many fields of physics'', as there are many examples and applications of PT symmetry in photonics in the review paper~\cite{ozdemir2019parity}. 

\begin{figure}[!htp]
    \subfigure[]{\includegraphics[width=0.4\textwidth]{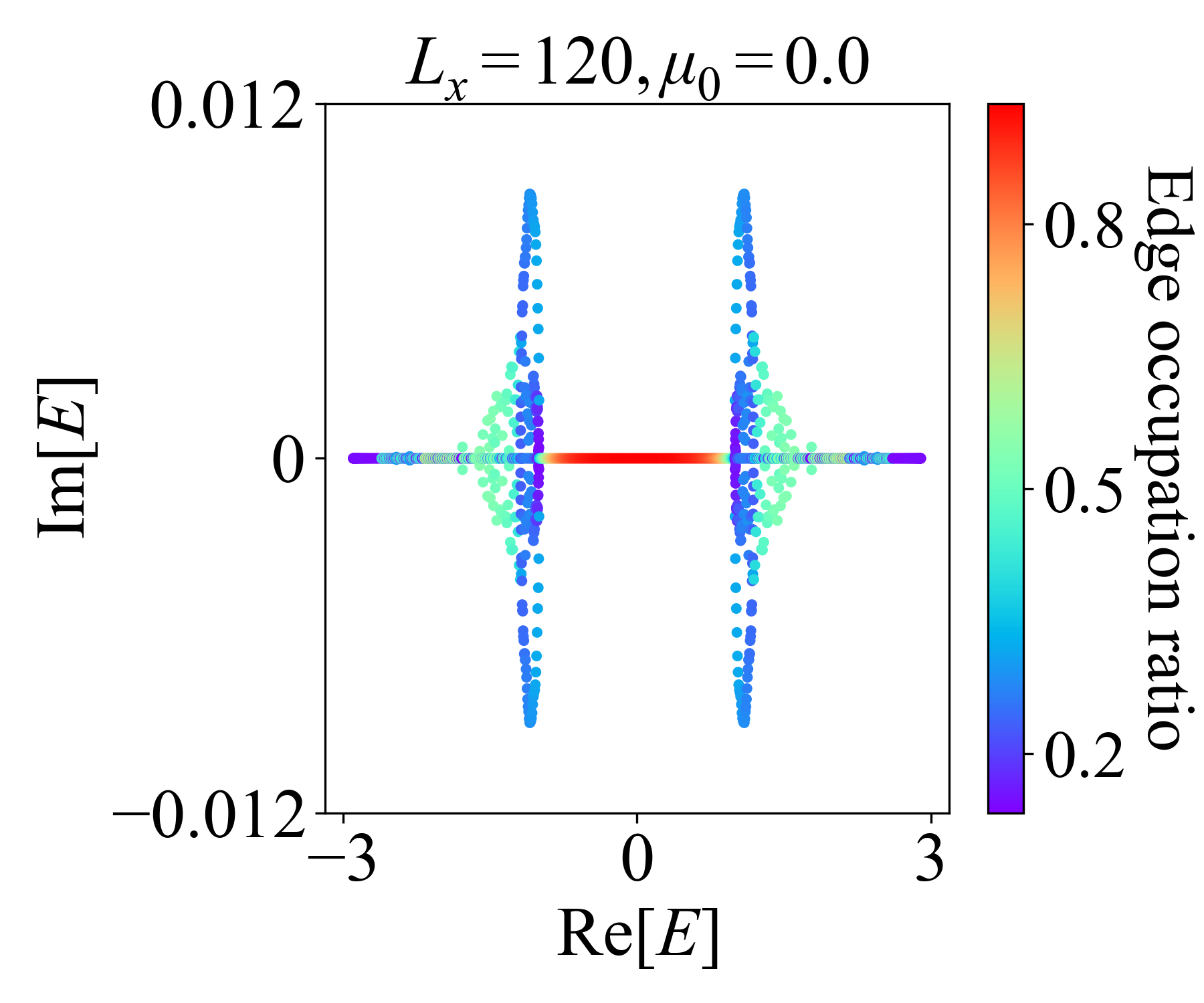}}
    \subfigure[]{\includegraphics[width=0.4\textwidth]{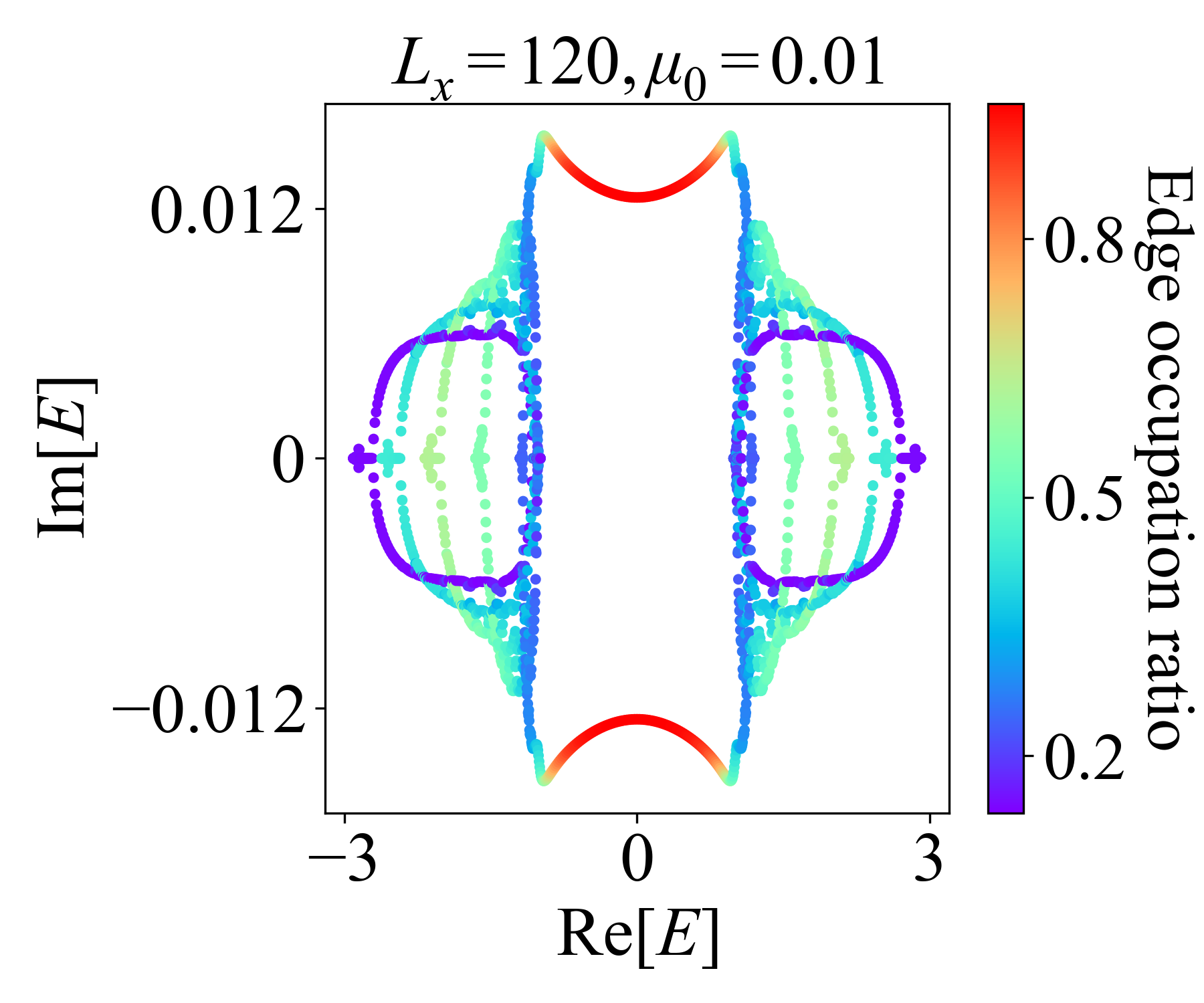}}
    \caption{\textcolor{black}{The spectrum of the bilayer system with (a) $\mu_0=0$ and (b) $\mu_0=0.01$. The parameters are $L_x=120, L_y=5, \gamma=0.02$.}}
    \label{rfig:bilayer_mu0}
    \end{figure}

However, as depicted in Fig.~\ref{rfig:bilayer_mu0}, \textbf{our model reveals that the topological spectrum of the bilayer system remains fully real when the coupling parameter $\mu_0$ is set to zero.} The transition from a real to a complex spectrum is induced by the introduction of the coupling $\mu_0$. While loss and gain (hopping asymmetry) do indeed break time-reversal symmetry in our model, they do not result in a net imaginary part associated with amplification, as illustrated in Fig.~\ref{rfig:bilayer_mu0}(a). It's only with the introduction of the coupling parameter $\mu_0$ that the topological spectrum becomes complex, as demonstrated in Fig.~\ref{rfig:bilayer_mu0}(b): the transition between these states is triggered by the coupling $\mu_0$. Here we take the same parameters as in Fig.~2 of the main text: $L_x=120, L_y=5, \gamma=0.02$. 

It is also interesting to note that the non-Hermitian Chern model alone can not induce the PT transition, as the topological spectrum remains real. It is the coupling between two oppositely NHSE layers that induces the PT transition.

\subsection{Comparison with the conventional PT transition mechanism}

Fundamentally, PT symmetry is broken when an evolving state begins to grow indefinitely ($\text{Im}(E) \neq 0$). As referee B correctly pointed out, previous literature (e.g., Ref.~\cite{ozdemir2019parity}) has established that whether the state grows infinitely or not is independent of the spatial characteristics of the system. Instead, it hinges solely on the magnitude of gain/loss. For instance, in a basic EP model, if $\gamma$ exceeds a certain threshold, PT symmetry breaks. 

However, what sets our paper apart is the introduction of a novel mechanism called topologically guided gain. We demonstrate that achieving a PT transition doesn't require tuning the $\gamma$ parameter. This is a significant departure from conventional mechanism. Instead, the  topologically guided gain, where the long-term behavior of a state is contingent only upon the width of spatial islands.

\section{Lattice hoppings in various physical systems}

In the main text, we used the term ``lattice hoppings'' to describe the coupling in the bilayer system. Here, we provide a more detailed explanation of the term.

In various physical systems,``lattice hoppings'' typically refers to the transfer or movement of particles or quasiparticles, such as electrons and photons, from one location to another. In experiments, the ``lattice hoppings'' can be realized in a variety of platforms, including photonics/optics proposals~\cite{chang2014parity,liu2018observation,ozdemir2019parity,yang2022concentrated}, quantum circuit proposals~\cite{lee2018topolectrical,zhang2020non,zou2021observation}, mechanical/acoustic proposals~\cite{helbig2020generalized,xiu2023synthetically}, and ultracold atomic proposals~\cite{gou2020tunable,li2020topological}.  This concept is central to understanding the behavior of these systems and can manifest differently depending on the context. 
\begin{itemize}
    \item In photonic systems, lattice hoppings can refer to the movement of photons between different optical modes within optical cavities or waveguides.
    \item In quantum circuits, lattice hoppings refer to the transfer of quantum spin states or quasiparticles between different qubits which are represented by say trapped ions or superconducting Josephson junction eigenstates.
    \item In mechanical or acoustic systems, lattice hoppings can be analogous to the transfer of vibrational energy or sound waves between different resonant modes of a mechanical structure.
    \item In electrical circuits, In electrical circuits arrays, a lattice hoppings term between two nodes corresponds to the admittance between them i.e. for instance, the ``lattice hoppings'' term between two capacitively coupled nodes is given by $i\omega C$. Hence ``lattice hoppings'' in an electrical circuit refer to how the voltages at various nodes are coupled.
    \item In systems of ultracold atoms, lattice hoppings refer to the orbital overlaps between nearby atoms.
\end{itemize}

\section{Physical understanding of the interlayer feedback}

In the main text, we largely provide mathematical understanding in our manuscript. We should have included more physical understanding. In the following, we first provide the physical understanding of the interlayer feedback. Then, we will discuss how the interlayer feedback can be realized in various platforms.

\subsection{Physical understanding of the interlayer feedback}

The interlayer feedback actually just arises from the interlayer coupling, which is denoted as $\mu_0$ in our manuscript. As discussed in the answer to your previous question regarding the PT transition, we claim that ``it is the couplings that induce such PT transition'', as visualized in Fig.~\ref{rfig:bilayer_mu0}. In our model, the interlayer feedback is a result of the coupling between two layers of Chern insulators with opposite NHSE directions. This coupling allows the states on one layer to influence the states on the other layer, and vice versa. The interlayer feedback is crucial for the topologically guided gain mechanism, as it enables the states on one layer to grow in magnitude due to the presence of states on the other layer. This feedback effect is what leads to the percolation-induced PT transition in our model.

\subsection{Realization of interlayer feedback in various platforms}

In general, coupling sites (on two layers) in different physical systems involves creating a connection that allows for the transfer or interaction between particles, quasiparticles, or fields at those sites. Below is a brief description of how interlayer feedback can be realized in various platforms.

\begin{itemize}
    \item In photonic systems, coupling between two sites, such as optical cavities or waveguides, can be achieved through evanescent wave coupling, where the mode fields of the waveguides overlap, allowing light to transfer from one to the other
    \item Mechanical coupling can be done by connecting two parts of a system with a component that can transmit vibrational energy or sound waves between them
    \item Electrical coupling often involves capacitors or inductors to couple the layers 
    \item In ultracold atomic systems, coupling between sites in an optical lattice can be controlled by dynamically changing the lattice parameters, such as the depth and spacing of the potential wells, to allow atoms to control the tunneling amplitude of the atoms
\end{itemize}

\section{Discussion on the velocity of the topological edge modes}

In the main text, we use the velocity as a constant, which results from the linear dispersion of the topological edge modes. 

Topological edge modes connect different bulk bands, and are generically linearly dispersing unless they are fine-tuned such that the linear gradient disappears and that the subleading quadratic dispersion dominates.

Below, firstly, we present an example of linear dispersion relations in topoelectrical circuits~\cite{hofmann2019chiral}, as we proposed a feasible approach to observe our results based on Ref.~\cite{hofmann2019chiral}, where the dispersion relations indicate a constant velocity. Then, we will address the fact that our primary concern is the averaged velocity, regardless of whether it remains constant.

\subsection{Example of linear dispersion relations in topological Chern insulators}

\begin{figure}[!htp]
    \centering
    \includegraphics[width=0.7\textwidth]{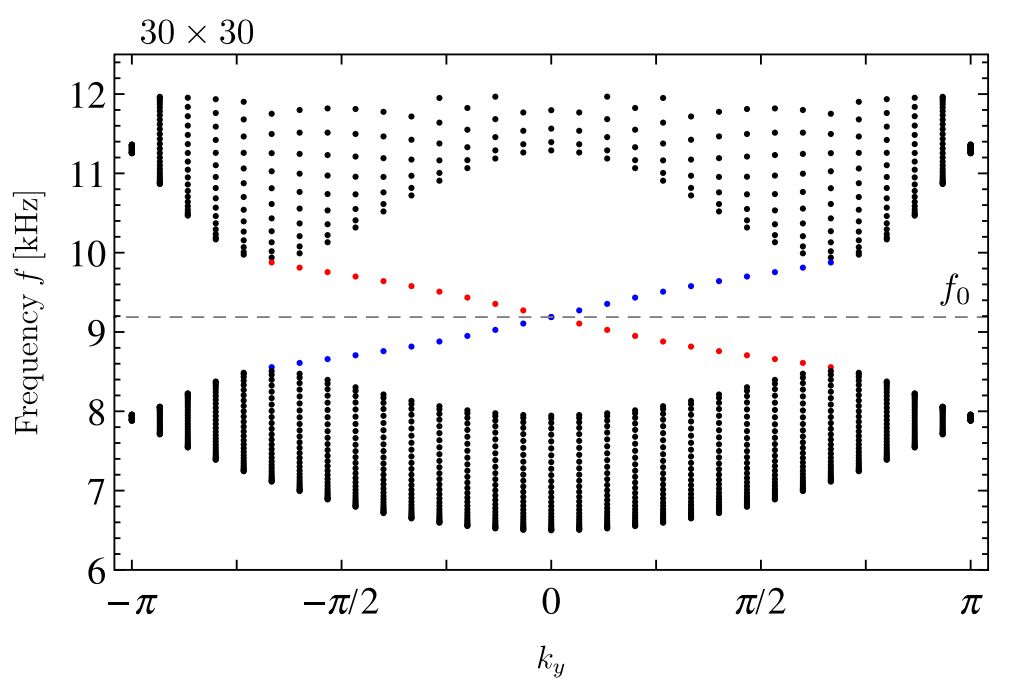}
    \caption{\textcolor{black}{Frequency band structure of the topological Chern insulators in the topoelectrical circuits, adapted from Ref.~\cite{hofmann2019chiral}. The linear dispersion relations in topological edge states indicate a constant velocity.}}
    \label{fig:dispersion}
    \end{figure}

In the previous response to the experimental porposal, we have introduced the topoelectrical
circuits model showing Chern insulator analogy, where the linear dispersion relations are observed. In Fig.~\ref{fig:dispersion}, we show the dispersion relations of the topological Chern insulators in the topoelectrical circuits, adapted from Ref.~\cite{hofmann2019chiral}. The linear dispersion relations indicate a constant velocity.

\subsection{Velocity does not necessarily need to be constant}

Furthermore, it is important to stress that our findings remain largely unchanged even if the velocity is not constant. The velocity of the edge modes is not pivotal in our research; our primary focus lies in topologically guided gain and percolation-induced PT symmetry breaking. While we utilize a ``constant'' velocity to elucidate this mechanism of the topologically guided gain, we could employ a more generalized averaged velocity. As long as the average velocity is not zero, the topologically guided gain will persist.


\end{document}